\documentclass[aps,11pt,nofootinbib]{revtex4-1}
\pdfoutput=1
\usepackage{graphicx}
\usepackage[margin=1in]{geometry}
\usepackage{amsmath}
\usepackage{amssymb}
\usepackage{color}
\usepackage[utf8]{inputenc}

\makeatletter
\renewcommand\@make@capt@title[2]{%
  \@ifx@empty\float@link{\@firstofone}{\expandafter\href\expandafter{\float@link}}%
   {\textbf{#1}}\@caption@fignum@sep#2\quad
}%
\makeatother

\newcommand{\mt}{\mathcal}

\begin{document}

\author{Sam Young$^1$}
\email{syoung@mpa-garching.mpg.de}
\author{Ilia Musco$^{2,3}$}
\email{iliamusco@icc.ub.edu}
\author{Christian T. Byrnes$^{4}$}
\email{C.Byrnes@sussex.ac.uk}

\affiliation{\\ 1) Max Planck Institute for Astrophysics, Karl-Schwarzschild-Strasse 1, 85748 Garching bei Muenchen, Germany} 

\affiliation{\\ 2) Institut de Ci\`encies del Cosmos, Universitat de Barcelona, \\Mart\'i i Franqu\`es 1, 08028 Barcelona, Spain} 

\affiliation{\\ 3) Laboratoire Univers et Th\'{e}ories, UMR 8102 CNRS, Observatoire de 
Paris, Universit\'{e} Paris Diderot, 5 Place Jules Janssen, F-92190 Meudon, France\\} 

\affiliation{\\ 4) Department of Physics and Astronomy, University of Sussex, Brighton BN1 9QH, United Kingdom\\}

\date{\today}

\title{Primordial black hole formation and abundance: contribution from the non-linear relation between the density and curvature perturbation}

\begin{abstract}
The formation and abundance of primordial black holes (PBHs) arising from the curvature perturbation $\zeta$ is studied. The non-linear relation between $\zeta$ and the density contrast $\delta$ means that, even when $\zeta$ has an exactly Gaussian distribution, significant non-Gaussianities affecting PBH formation must be considered. Numerical simulations are used to investigate the critical value and the mass of PBHs which form, and peaks theory is used to calculate the mass fraction of the universe collapsing to form PBHs at the time of formation. A formalism to calculate the total present day PBH abundance and mass function is also derived. It is found that the abundance of PBHs is very sensitive to the non-linear effects, and that the power spectrum $\mathcal{P}_\zeta$ must be a factor of ${\mathcal O} (2)$ larger to produce the same number of PBHs as if using the linear relation between $\zeta$ and $\delta$ (where the exact value depends on the critical value for a region to collapse and form a PBH). This also means that the derived constraints on the small-scale power spectrum from constraints on the abundance of PBHs are weaker by the same factor.
\end{abstract}

\maketitle

\tableofcontents

\section{Introduction}

Primordial black holes (PBHs) could be formed from the gravitational collapse of large curvature perturbations created during 
cosmological inflation shortly after re-entering the cosmological horizon
at early times \cite{Hawking:1971ei,Carr:1974nx,Carr:1975}. If a density perturbation is above a threshold $\delta_c$, then an apparent horizon will form 
during the collapse, otherwise it will quickly disperse into the surrounding local environment. The mass of the resulting 
PBH is strongly related to the scale and amplitude of the perturbation from which it formed, with more massive black holes forming 
from larger-scale perturbations which enter the horizon at a later time. PBHs can theoretically form with any mass, and can provide 
a natural explanation for any observed black holes with masses which are not easily explained by the standard picture of black hole 
formation from collapsing stars. PBHs which formed with a mass below $~10^{15}g$ would have 
evaporated by today (ignoring the possible 
accretion after formation), but more massive PBHs would persist into the present epoch.

PBHs still represent a viable dark matter candidate, although there are numerous constraints on the abundance of PBHs of varying 
masses (see \cite{Carr:2017jsz,Kuhnel:2017pwq,Bellomo:2017zsr} for recent discussions of the constraints for a broad mass spectrum), and clusters of PBHs could explain the early formation of 
super-massive black holes found in the centres of galaxies. In recent years, there have been several interesting observations which may 
hint towards the existence of PBHs \cite{Clesse:2017bsw}. For a review of PBH formation and constraints, see \cite{Green:2014faa,Sasaki:2018dmp}.


There have been many attempts to detect them by their indirect effects on the universe. Ignoring the possible observations described
above, no positive detection has been made. 
However, the non-detection of PBHs constrains their abundance.
The abundance of PBHs is typically stated in terms of $\beta$, the energy fraction of the universe which goes into PBHs at the time of 
formation. The abundance of small PBHs ($<10^{15}g$) in the early universe which would have evaporated by today can be constrained
by looking for the effects of the radiation from their evaporation, whilst the abundance of more massive PBHs ($>10^{15}g$) can be 
constrained by their gravitational effects. Because PBHs of different masses form from different scale perturbations, the constraints on 
different mass PBHs can be used to place a constraint on different scales of the primordial power spectrum in the early universe - though 
these constraints are sensitive to primordial non-Gaussianity in the early universe. Because PBHs form on scales much smaller than those 
observable in the cosmic microwave background (CMB) or large-scale structure of the universe (LSS), they therefore place the only available constraints on the small-scale power spectrum - and can be used to constrain models of inflation. In order for an interesting number of PBHs 
to form however, the power spectrum must become orders of magnitude larger on small scales than is observed in the CMB or LSS 
($\mathcal{O}(10^{-2})$ compared to $\mathcal{O}(10^{-9}$)). Therefore the derived constraints on the power spectrum are much weaker, but they span a much larger range of scales, including scales which cannot be probed by any other method \cite{Byrnes:2018txb}.

There are many different models for cosmological inflation which would predict a large number of PBHs to form in the early universe. 
For example, the running mass model \cite{Drees:2011hb}, axion inflation \cite{Bugaev:2013fya,Ozsoy:2018flq}, or a waterfall transition during hybrid 
inflation \cite{GarciaBellido:1996qt,Lyth:2012yp,Bugaev:2011wy}, a quartic action during inflation or a variable sound speed \cite{Ballesteros:2018wlw}, amongst many others. A metric perturbation in the form of the curvature perturbation $\zeta$ is 
typically used to study cosmological perturbations generated with the different models and to predict their observable consequences\footnote{Often
 $\mathcal{R}$ is used instead to denote the curvature perturbation, whilst $\zeta$ is used to describe the curvature perturbation only on a uniform density slicing.}. The curvature perturbation $\zeta$ appears in the metric in the comoving uniform-density gauge as
\begin{equation} \label{metric}
\mathrm{d} s^2 =-\mathrm{d}\eta^2 + a^2(t)e^{2\zeta}\mathrm{d}X^2,
\end{equation}
where $\eta$ is the conformal time, $a(t)$ is the scale factor and $X$ represents the three comoving spatial coordinates. In order to translate the constraints on PBH 
abundance into constraints on models of inflation (or alternatively to predict PBH abundances from a given model) it is desirable to relate the 
primordial curvature perturbation power spectrum $\mathcal{P}_\zeta$ to the abundance of PBHs of different masses.

The formation of PBHs from a non linear metric perturbation was initially studied by Shibata and Sasaki \cite{Shibata:1999zs}, 
which was then used to derive a relation between the abundance of PBHs and the power spectrum $\mathcal{P}_\zeta$ \cite{Green:2004wb}.
Around the same time Niemeyer and Jedamzik performed a numerical study of PBH formation using instead an initial perturbation of 
the energy density \cite{Niemeyer:1997mt,Niemeyer:1999ak}. For a long time the abundance of PBHs was calculated assuming that regions where the curvature perturbation $\zeta$ was above a critical value $\zeta_c$ of order unity. However, it has since been shown that the curvature perturbation $\zeta$ is not a suitable parameter 
to use to determine whether a region will form a PBH or not, due to the effect of super-horizon modes on the calculation, and that the density 
contrast should be used instead \cite{Young:2014ana}. The effect of super-horizon modes on PBH formation is discussed in detail in \cite{Harada:2015yda}. Nonetheless, in the following years it has been typical to use the curvature perturbation directly to calculate the abundance - which is 
valid in the case that an approximation is being used (as described in \cite{Young:2014ana}) or in the case that a narrow peak in the power spectrum 
is being considered (so that large perturbations only exist at one scale). Papers which have used the density contrast $\delta\rho/\rho_b$ rather than the curvature perturbation for the calculation of the abundance have since used a linear relation between the two parameters (as in the recent paper \cite{Germani:2018jgr} for example),
\begin{equation}
\frac{\delta\rho}{\rho_b} = -\frac{2(1+\omega)}{5+3\omega} \left(\frac{1}{aH}\right)^{2} \nabla^2\zeta,
\end{equation}
where $\omega=1/3$ is the equation of state parameter during the radiation dominated epoch of the early universe, $\rho_b$ is the background density and 
$(aH)^{-1}$ is the comoving cosmological horizon scale. However, this 
expression ignores the non-linear relation between the curvature and the energy density profile. Starting for the first time from simulations of PBH formation arising from perturbations in the curvature perturbation $\zeta$, the aims of this paper are to investigate how the fully non-linear relation between $\zeta$ and $\delta\rho/\rho_b$ affects the calculation of the abundance of primordial black holes, and to derive the most accurate relation to date between the primordial curvature perturbation power spectrum $\mathcal{P}_\zeta$ and the abundance of PBHs at formation $\beta$.

We note that reference \cite{Yoo:2018esr} introduced a new calculation which included the non-linear effect. The abundance of PBHs was calculated using a method based on estimating the critical heights of peaks in the curvature perturbations $\zeta$, and then calculating the abundance of PBHs by utilising the Gaussianity of $\zeta$. Using the critical height of peaks in $\zeta$ is only valid for a narrow power spectrum, such that perturbations exist only on one scale (as they note in section 4), whereas the method presented here can be applied to power spectra of any width. Reference \cite{Yoo:2018esr} stated that the abundance of PBHs they calculated was greatly \emph{increased} compared to some previous calculations, whereas it is shown here that considering the non-linear relation \emph{decreases} the abundance compared to the linear calculation. The apparent contradiction in conclusions can be attributed to differences in what is considered as the ``standard calculation'' between that paper and ours. We compare our results using the same method from using the linear or non-linear relations between $\delta$ and $\zeta$, whereas \cite{Yoo:2018esr} compares very different methods.

The paper is organised as follows: in section \ref{sec:ic} we will discuss the set up of the initial conditions in the density contrast $\delta\rho/\rho_b$ arising from an initial curvature perturbation $\zeta$. In section \ref{sec:collapse} we will discuss the criteria which should be used to determine whether an initial perturbation will eventually collapse to form a PBH. In section \ref{sec:results} we will discuss the simulation procedure used and the numerical results obtained from the simulations.  Finally, in section \ref{sec:abundance} we will show how the abundance of PBHs $\beta$ can be obtained from the curvature perturbation power spectrum. Our findings are summarised in section \ref{sec:discussion}, leaving some details of our calculations to an appendix.
 
\section{Cosmological perturbations in the super horizon regime}\label{sec:ic}

In this section we will first describe the general relation between the curvature perturbation $\zeta$ and the density contrast $\delta\rho/\rho_b$ 
before analysing a specific parametrization of $\zeta$ that allows us to vary the profile of $\delta\rho/\rho_b$. This allows us, with the help of 
numerical simulations (see Section \ref{sec:results}), to span almost all the possible range of values of $\delta_c$. Throughout this 
paper, we assume that perturbations large enough to form PBHs are spherically symmetric. This is justified because such peaks 
must be extremely rare \cite{Bardeen:1985tr}, and the perturbation profile is therefore defined using only a radial coordinate $r$.

Note that the curvature perturbation $\zeta$ in the literature is typically defined on a uniform density slicing, whilst the density contrast 
$\delta\rho/\rho_b$ is defined on a constant cosmic time slicing. However, in the super-horizon regime described in the following section the 
difference between these two gauges is a higher-order correction which can be neglected (see \cite{Lyth:2004gb} and the references therein).

\subsection{Gradient expansion approach}
In the super-horizon regime perturbations have a length scale much larger than the cosmological Hubble horizon. In this regime it is possible to have an analytic treatment,
usually called the gradient expansion or long-wavelength approximation \cite{Shibata:1999zs,Salopek:1990jq}, based on expanding 
the exact solution in a power series of a small parameter ($\epsilon\ll1$) that is conveniently identified with the ratio between the 
Hubble radius $1/H(t)$ and the length scale $L$ characterising the perturbation
\begin{equation}
\epsilon(t) \equiv \frac{1}{H(t)L},
\label{epsilon_def}
\end{equation}   
where a particular value of $\epsilon$ corresponds to a particular value of the time $t$.  

When $\epsilon\ll1$ the curvature profile $\zeta(r)$ 
is conserved (time independent) and each spatial gradient is multiplied by $\epsilon$, expanding the equations in a power series in 
$\epsilon$ up to the first non-zero order. Putting $\epsilon=1$, which defines the horizon crossing time, one obtains the spatial dependence 
of the different matter/geometrical variables, related to the right/left hand side of the Einstein equations, in terms of the conserved curvature 
profile $\zeta(r)$. The curvature profile represents the fundamental source of the perturbation, embedded into the metric \eqref{metric} from 
the asymptotic limit, $t\to0$.  Although this approach reproduces the time evolution of linear perturbation theory when $\epsilon\ll1$, it also 
allows one to consider non-linear curvature perturbations if the spacetime is sufficiently smooth on scales greater than $L$ \cite{Lyth:2004gb}. 
This is equivalent to pressure gradients being a higher-order correction in $\epsilon$, which corresponds to a self-similar growth of the 
perturbation, conserving the spatial profile. 
 
In the gradient expansion the non-linear relation between the density contrast $\delta\rho/\rho_b$ defined on a comoving uniform-cosmic time slicing 
and the curvature perturbation $\zeta$ is given by \cite{Harada:2015yda,Musco:2018rwt,Yoo:2018esr}, which is exact up to the first non-zero terms in $\epsilon$
\begin{equation} 
\frac{\delta\rho}{\rho_b}(r,t) = - \frac{4(1+\omega)}{5+3\omega} \left(\frac{1}{aH}\right)^2 e^{-5\zeta(r)/2} \nabla^2 e^{\zeta(r)/2},
\label{eqn:non-linear}
\end{equation}
where $\omega\equiv p/\rho$ is the equation of state parameter, and 
\[ \nabla^2 = \frac{\rm d^2}{\rm dr^2}  + \frac{2}{r} \frac{\rm d}{\rm dr} \]
is the Laplacian written assuming spherical symmetry. 
There are two non-linear effects contained within equation (\ref{eqn:non-linear}): the exponential term $\exp(-2\zeta(r))$ and the quadratic term 
of the first derivative $\left( \zeta'(r)\right)^2$, where the prime denotes a derivative with respect to the radial coordinate $r$. Because PBHs form 
from large perturbations, the effect of the non-linear components is comparable with the linear term and should not be neglected. 



As explained in \cite{Musco:2018rwt}, whether a cosmological perturbation is able to form a PBH depends on the amplitude measured at the 
peak of the compaction function defined as
\begin{equation}
\label{a}
\mathcal{C}(r,t) \equiv 2\frac{M(r,t)-M_b(r,t)}{R(r,t)} \,,
\end{equation}
where $R(r,t)$ is the areal radius, $M(r,t)$ is the Misner-Sharp mass within a sphere of the radius $R$ and  $M_b(r,t)=4\pi \rho_b(r,t)R^3(r,t)/3$ 
is the background mass within the same areal radius calculated with respect to a Friedmann-Lema\^{i}tre-Robertson-Walker (FLRW) universe. In the superhorizon regime, applying the gradient 
expansion approximation, the compaction function is conserved and is 
related to $\zeta(r)$ as \cite{Harada:2015yda,Musco:2018rwt,Yoo:2018esr}
\begin{equation} \label{compact}
\mathcal{C}(r) = - f(w) r\zeta'(r) \left[ 2+r\zeta'(r) \right] \,,\quad \quad f(\omega) = \frac{3(1+\omega)}{5+3\omega} \,.
\end{equation} 
We can then compute the length-scale of the perturbation, identified as the location $r_m$ where the compaction function is maximized 
($\mathcal{C}'(r_m) = 0$), which gives
\begin{equation} \label{eq_rm}
\zeta'(r_m)+r_m\zeta''(r_m)=0 \,.
\end{equation}

Measuring the curvature perturbation with $\zeta(r)$ introduces an intrinsic rescaling of the comoving coordinate with respect to the background 
FLRW solution, because the exponential term appearing in the metric \eqref{metric} introduces a local perturbation of the scale factor which 
depends on the local value of the curvature. This implies that the horizon crossing time $t_H$ is defined in real space when
\begin{equation} \label{hor_crossing}
a(t_H) H(t_H) r_m e^{\zeta(r_m)} = 1, 
\end{equation}
and therefore according to the definition of $\epsilon$ given above, the physical length scale of the perturbation, to be called $R_m$ from here 
onwards, is given by 
\begin{equation}
R_m(r,t) = a(t) r_m e^{\zeta(r_m)} \,.
\end{equation}

The perturbation amplitude can be measured as the mass excess of the energy density within the scale $R_m$, measured at the horizon crossing 
time $t_H$. Although in this regime the gradient expansion approximation is not very accurate, this represent a well defined criterion that allows a 
consistent comparison between the amplitude of different perturbations (see \cite{Musco:2018rwt} for more details). Computing the mass excess as 
the integral of the density contrast averaged over the background volume $V_{R_m} = \frac{4\pi}{3}R_m^3$, the amplitude of the perturbation is given by
\begin{equation}
\label{delta}
\delta(r_m,t_H) \equiv \frac{3}{R^3_m} \int\limits_0^{R_m} \, \frac{\delta\rho}{\rho_b} R^2 {\rm d} R = 
\frac{3}{(r_m e^{\zeta(r_m)})^3} \int\limits_0^{r_m} \, \frac{\delta\rho}{\rho_b}(r,t_H) (re^{\zeta(r)})^2 (re^{\zeta(r)})' {\rm d}r  \,,
\end{equation}
and using the explicit expression of $\delta\rho(r,t)/\rho_b$ in terms of the curvature profile seen in \eqref{eqn:non-linear}, we get 
\mbox{$\delta_m \equiv \delta(r_m,t_H) = \mathcal{C}(r_m)$}, which satisfies the fundamental relation  \cite{Musco:2018rwt}
\begin{equation}
\delta_m = 3 \frac{\delta\rho}{\rho_b}(r_m,t_H)  
\label{delta_m}
\end{equation}
for any curvature profile, $\zeta(r)$. 

\subsection{Initial conditions}
We will now study the main feature of the density profile when an explicit parameterization of the curvature profile $\zeta(r)$ is 
specified as
\begin{equation}
\zeta_\alpha(r) = \mathcal{A} \exp\left[- \left(\frac{r}{r_m}\right)^{2\alpha}\right] ,
\label{zeta_Gauss}
\end{equation}
where $\mathcal{A}$ and $r_m$ respectively denote the amplitude and the scale of the 
perturbation. Inserting this into \eqref{eqn:non-linear}, the corresponding density profile is given by
\begin{equation}
\frac{\delta\rho_\alpha}{\rho_b} =  \left(\frac{1}{aH}\right)^2 \frac{4}{3}f(\omega) \alpha \left(\frac{r}{r_m}\right)^{2\alpha}
\left[ (2\alpha+1) - \alpha\left(\frac{r}{r_m}\right)^{2\alpha} \left( 2 + \zeta_\alpha(r) \right) \right] 
\frac{\zeta_\alpha(r)}{r^2e^{2\zeta_\alpha(r)}} \,,
\label{delta_zeta}
\end{equation}
and then inserting \eqref{zeta_Gauss} into \eqref{compact} and \eqref{eq_rm}, we can calculate the corresponding perturbation 
amplitude
\begin{equation}
- r_m \zeta'_\alpha(r_m ) =  \frac{2\mathcal{A}\alpha}{e}  \quad 
\Rightarrow \quad \delta_m = 4 f(\omega)\frac{\mt{A}\alpha}{e} \left( 1 - \frac{\mt{A}\alpha}{e} \right),
\label{delta_mz}
\end{equation} 
which gives the value of $\mathcal{A}$ in terms of the averaged amplitude $\delta_m$
\begin{equation} \label{zeta_peak}
\mt{A} = \frac{e}{2\alpha} \left( 1 - \sqrt{ 1 - \frac{\delta_m}{f(\omega)} } \,\right). 
\end{equation}
This behaviour of $\delta_m$ in terms of $\mt{A}$ also shows that there is a maximum value of $\delta_m=2/3$ corresponding to 
$\mt{A} = e\alpha/2$, that represents the transition between PBHs of type I and type II, after which formation of PBHs cannot be studied 
in terms of $\delta_m$ but only in terms of $\zeta$ \cite{Kopp:2010sh}.  

\begin{figure*}
\vspace{-1.5cm}
  \includegraphics[width=0.49\textwidth]{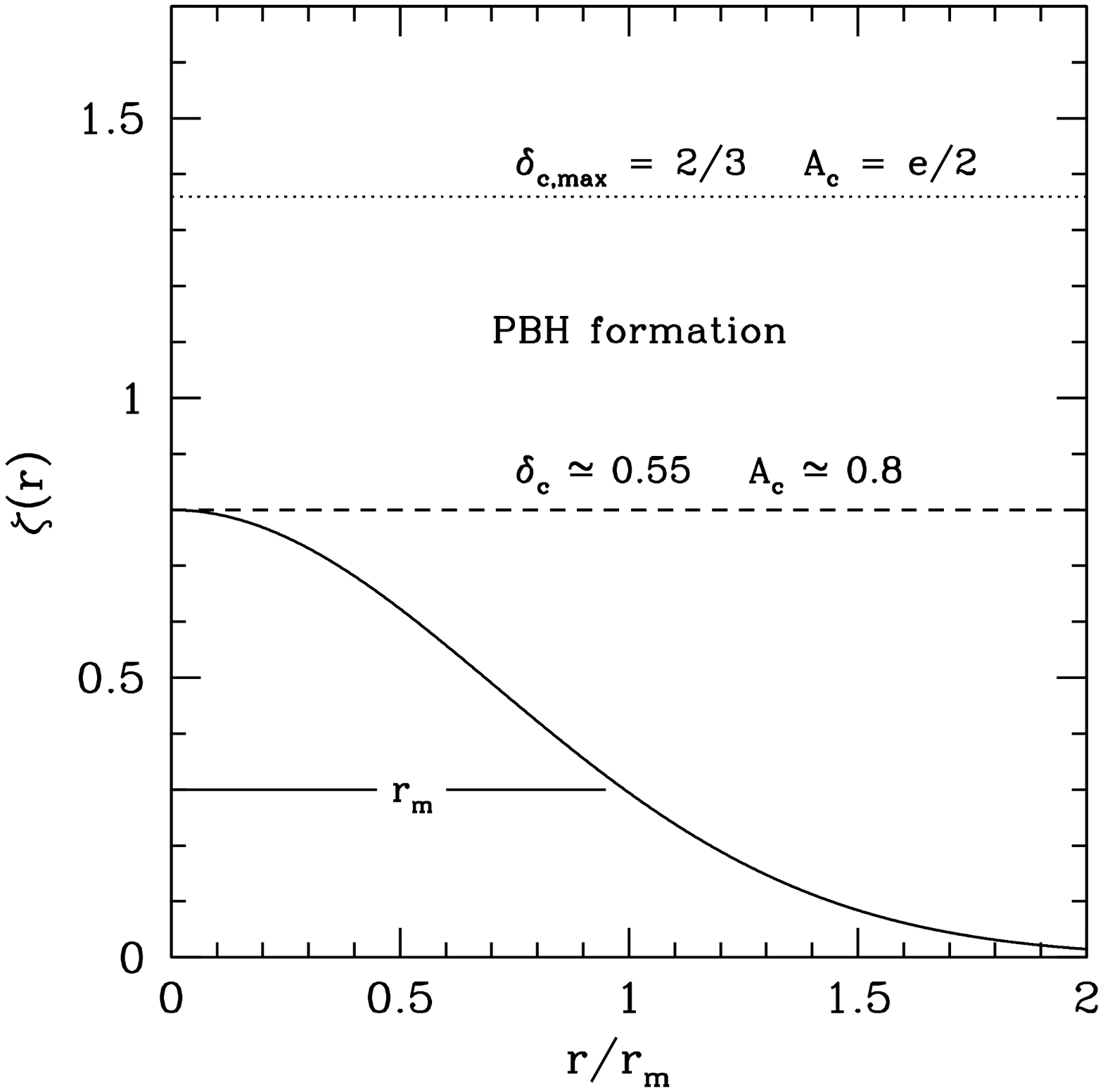} 
  \includegraphics[width=0.49\textwidth]{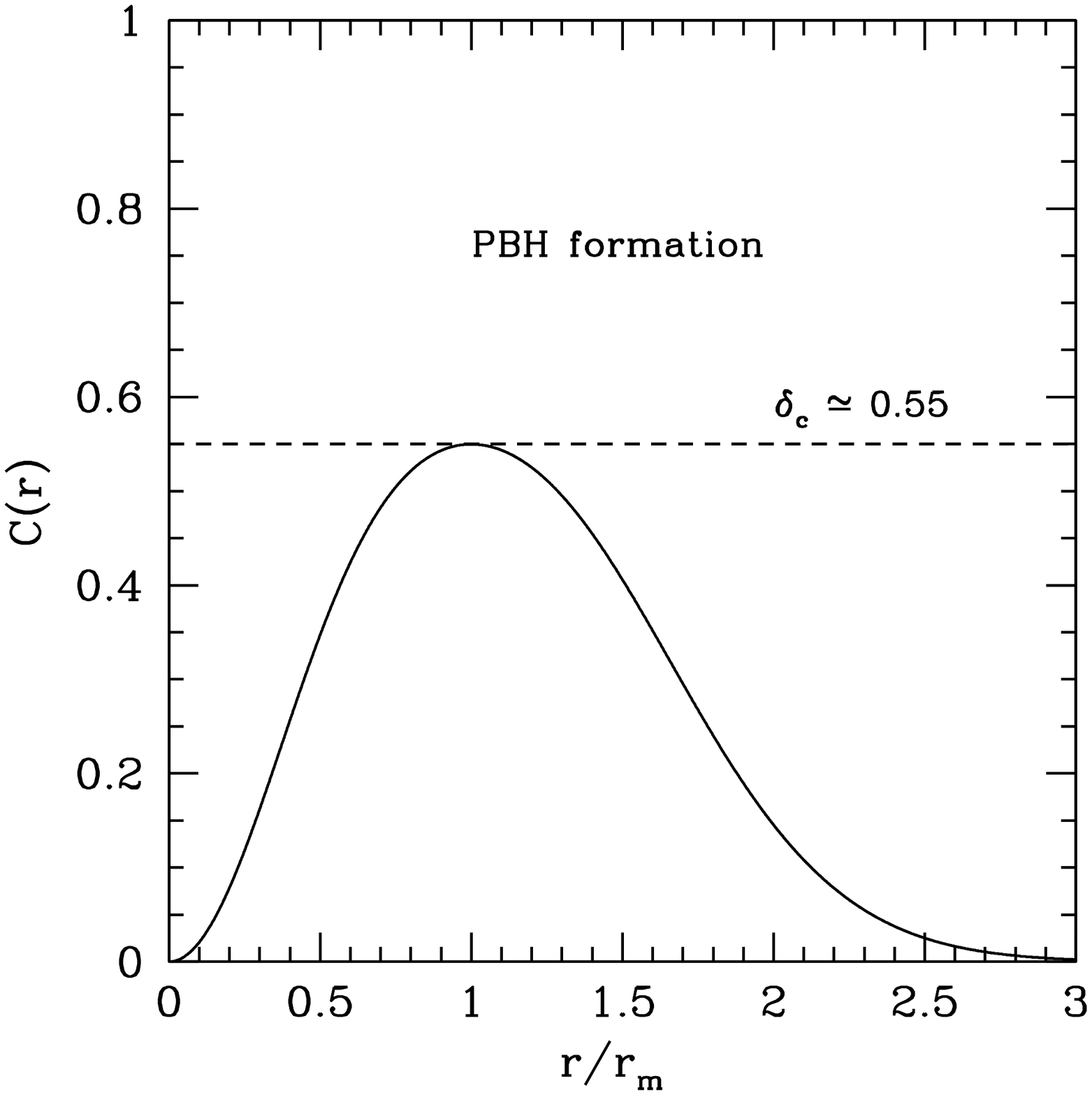} 
  \vspace{-2.0cm}

  \caption{ The left plots shows the critical curvature profiles given by \eqref{zeta_Gauss} for $\alpha=1$ ($\mt{A}_c\simeq0.8$ and 
  $\delta_c\simeq0.55$), while the right plot shows the corresponding ${\cal C}(r)$ profile. The upper limit $\delta_{c,max}=2/3$ 
  represents the maximum theoretical upper limit of $\delta_c$.}
  \label{zeta_r}
\end{figure*}

Using numerical simulations we have calculated the critical values for PBHs using initial condition in terms of 
$\zeta(r)$ given by \eqref{zeta_Gauss}, finding that  the value of $\delta_c$ is varying in terms of $\alpha$ (see 
 section \ref{sec:results} for more details). In the particular case of $\alpha=1$ we get $\delta_c\simeq0.55$ 
which corresponds to a value of \mbox{$\mt{A}\simeq0.80$}. In the left frame of figure \ref{zeta_r} we have plotted 
the critical profile of $\zeta(r)$ as function of $r/r_m$ identifying the critical peak $\mt{A}_c$ of the profile. In this plot 
we are also showing the value of $\mathcal{A}$ corresponding to the maximum limit of $\delta_{c,max}=2/3$. In the right hand 
plot of figure \ref{zeta_r} we show the corresponding compaction function $\mathcal{C}(r)$ with the peak amplitude of $\mathcal{C}(r_m)$ being equal to $\delta_c$.  

\begin{figure*}[t!]
\vspace{-1.5cm}
\centering
 \includegraphics[width=0.49\textwidth]{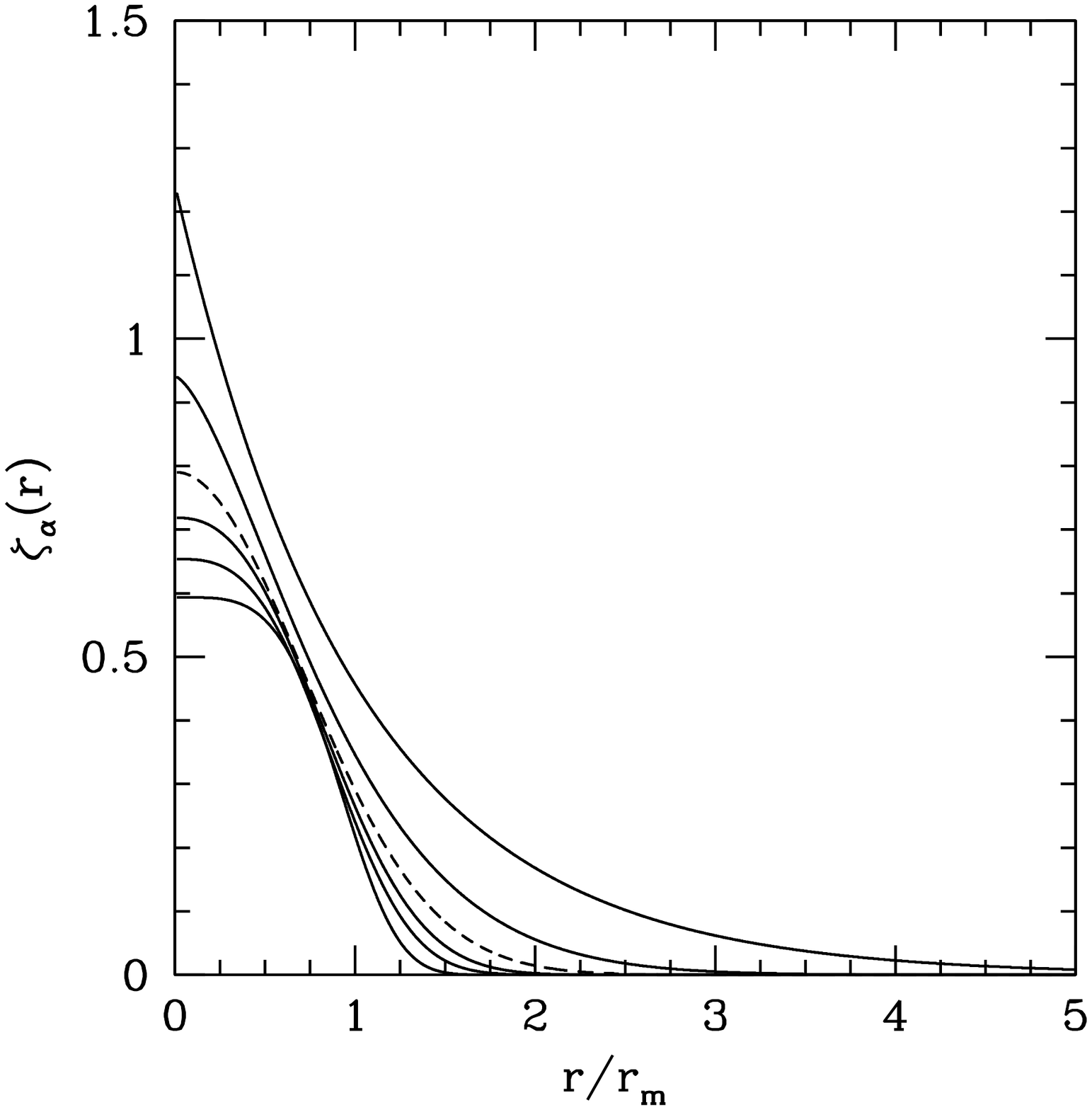} 
  \includegraphics[width=0.49\textwidth]{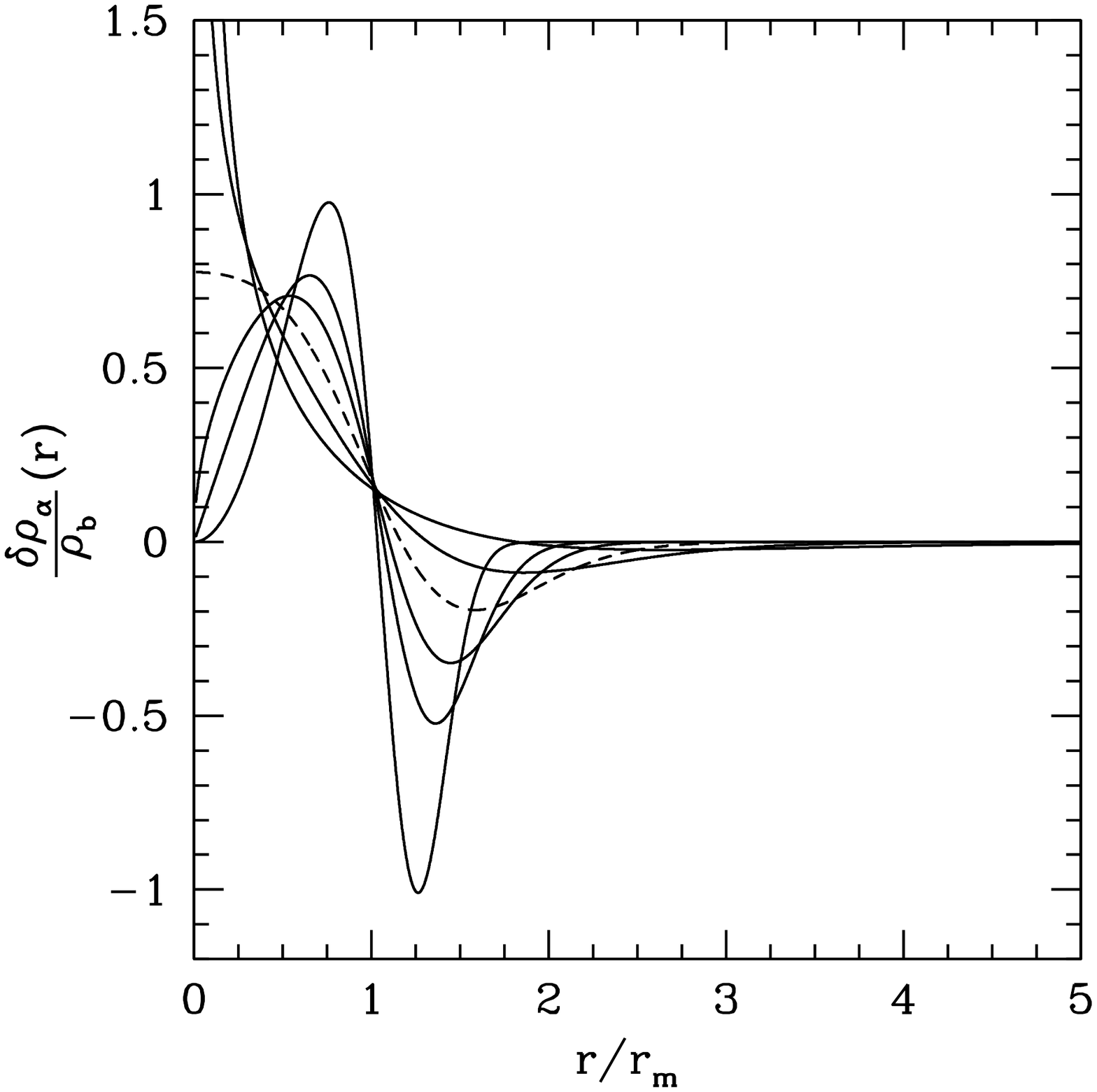} 
  \vspace{-2.0cm}

  \caption{ The left plot shows the critical profiles $\zeta_\alpha(r)$ given by \eqref{delta_zeta} plotted against 
  $r/r_m$ for \mbox{$\alpha=0.5, 0.75, 1.0, 1.25, 1.50, 2.0$} where for increasing values of $\alpha$
  the central value $\mathcal{A}_c$ is decreasing. The right plot shows the corresponding critical profiles of
  $\delta\rho_\alpha(r)/\rho_b$ given by equation \eqref{delta_zeta} plotted against $r/r_m$: for $\alpha<1$ the profile
  has a spiky shape, with increasing steepness for decreasing value of $\alpha$; for $\alpha=1$ the profiles is 
  centrally peaked with $(\delta\rho)'/\rho_b \to 0$ when $r\to0$ while for $\alpha>1$ the profile has an increasing off-peak. 
  In both plots the dashed line corresponds to the smooth centrally peaked case with $\alpha=1$.}

  \label{delta_rho}
\end{figure*}

In figure \ref{delta_rho} we plot the $\zeta$-profiles (plotted in the left panel) and the corresponding density contrast (right panel) 
measured at horizon crossing, defined by \eqref{hor_crossing}, for the threshold values $\delta_c$ associated to each shape. 
Although the $\zeta$-profiles given by \eqref{zeta_Gauss} 
are always centrally peaked, the energy 
density profile is centrally peaked only for $\alpha\leq1$. In particular, only the case of $\alpha=1$, 
plotted with a dashed line, is smooth at $r=0$, giving a finite value of the peak of the density contrast, while for smaller values 
of $\alpha$ the peak is diverging. Nevertheless the amplitude of the perturbation, measured by the averaged value $\delta_m$ 
always remains finite.

The right panel of figure \ref{delta_rho} also shows that for $\alpha>1$ the density contrast is off-centred, with  an increasing value 
of the peak for larger values of $\alpha$. One can see therefore that a $\zeta$-profile with a centrally finite peak does not always 
corresponds to the same type of peaks in the density contrast, because of the non-linear expression given by \eqref{eqn:non-linear} 
and the correspondence among the peaks is guaranteed only at the linear level. In general, the correspondence between the peaks in 
$\zeta$ and the peaks in $\delta\rho/\rho_b$ requires the assumption that $\zeta'(r)=0$ at $r=0$, such that in the centre the only non-linear 
effect is given by the exponential term $\exp{(-2\zeta(r))}$ which reduces the amplitude. 

A second issue already mentioned is that finite peaks of $\zeta$ do not always correspond to finite values of the peak of the 
density contrast, which happens here for $\alpha<1$. This is because the $\zeta$-profiles for $\alpha<1$ are not smooth in the 
centre (they are not infinitely differentiable), and the term $\zeta'(r)/r$ diverges in the limit $r\to0$.  Such peaks are of course 
unphysical and this divergence can be removed with a transfer function or a smoothing function which removes the small 
scale power, but there is lack of knowledge in the literature about which is the correct form of the non-linear transfer function 
to be used for a radiation fluid. Note that the transfer function should always be taken into account, because strictly 
speaking the curvature is exactly conserved only for $\epsilon\to0$. Because in practice a finite value of $\epsilon$, corresponding to 
a finite initial time $t_i$, needs to be chosen, the effects of the pressure gradients within a region of the size of the initial sound horizon, 
which in radiation is  $R_s = (\sqrt{3}H)^{-1}$, are not completely negligible even at the initial time. For spiky shapes (which have $\zeta'(r)/r\neq0$  in the centre), like those obtained from \eqref{zeta_Gauss} with $\alpha<1$ and the effect of the transfer function 
might significantly change the amplitude of the peak, while the value of the averaged $\delta_m$ hardly changes.

Because for every $\zeta$-peak with a finite amplitude there is always a peak of the compaction function ${\cal C}$ evaluated at 
$r_m$, with finite amplitude equal to $\delta_m$, we have used the averaged amplitude $\delta_m$ to calculate the 
abundance of PBHs (see section \ref{sec:abundance}), using the linear transfer function to correct the value of the peak of 
the density contrast at $r=0$, leaving a study of the effects of the non-linear transfer function to future work. 

\section{Criterion for collapse}
\label{sec:collapse}

In order for a perturbation to collapse into a PBH, the density must exceed some critical threshold. The original work by Carr 
\cite{Carr:1975}, using a Jeans length argument, provided an order of magnitude estimate  for the 
threshold  $\delta_c \sim \omega=1/3$ at horizon crossing for a radiation fluid. Since then, there has been 
extensive work to determine the collapse threshold \cite{Shibata:1999zs,Niemeyer:1999ak,Musco:2004ak,Polnarev:2006aa,Musco:2012au,Nakama:2013ica,Harada:2015yda,Musco:2018rwt}, as well as discussions about the appropriate parameter to use to determine whether a perturbation will collapse \cite{Green:2004wb,Young:2014ana,Germani:2018jgr,Yoo:2018esr}. The collapse threshold is typically obtained from simulating the evolution of a perturbation as it reenters the cosmological horizon, although analytic attempts have been made, 
neglecting the effects of pressure gradients \cite{Harada:2013epa}. 

As discussed previously, in order to determine a clear criterion to distinguish which perturbations are able to form a PBH, the density contrast $\delta\rho/\rho_b$ should be used rather than a metric perturbation such as the curvature perturbation $\zeta$. There has been much ambiguity in the literature about how this critical amplitude is calculated and used (especially between the different communities of relativists modelling PBH formation and cosmologists calculating the abundance of PBHs).  It is the aim of this section of the paper to discuss how this should be defined. Spherical symmetry is typically assumed when modelling PBH formation, again justified by the fact that such peaks are large and rare \cite{Bardeen:1985tr} - although non-spherical symmetry have been considered \cite{Harada:2015ewt}. In this section, we will discuss several ambiguities within the literature over how the criteria for collapse should be defined\footnote{A full description and investigation of these considerations is beyond the scope of this paper, although is discussed further in \cite{Young:2019osy}.}:

\begin{itemize}

\item {\bf The time at which PBH abundance should be calculated.} The threshold for collapse is normally stated in terms of the time-independent component of the density contrast, during the linear regime whilst a perturbation is super-horizon \cite{Musco:2018rwt}. In the linear regime, the density contrast grows proportionally to the parameter $\epsilon$ (equation \eqref{epsilon_def}), which is the ratio of the perturbation scale to the horizon scale at a given scale. Taking the time-independent component is therefore equivalent to setting this parameter to unity, which has resulted in many papers treating this as the value of the density contrast at horizon crossing. 
Ideally, the abundance of PBHs should be calculated by considering the perturbations on super-horizon scales, long before they reenter the horizon.

\item {\bf Should the peak value of the density contrast be used, or the smoothed density contrast?} The peak value at the centre of a density perturbation was used in a recent paper \cite{Germani:2018jgr} to determine the abundance of PBHs. This is valid when the distribution is already smooth on scales smaller than the scale being considered (as was considered in that paper), which is generally not the case unless a power spectrum with a very narrow peak is being considered. In order to investigate PBH formation over a wider range of scales, it is necessary to use a smoothing function. We also consider the fact that (for perturbations of arbitrary profile shapes), the threshold value for collapse of the central peak varies from $2/3$ to infinity, whilst the critical value for collapse of the top-hat smoothed density contrast varies from $~0.41$ to 2/3 - a much smaller range of values. 
It was also discussed in section \ref{sec:ic} that, for certain $\zeta$-profiles, the peak in the density contrast may be off-centred\footnote{It was shown in \cite{Musco:2018rwt} that off-centered profiles evolve to have a central peak, with almost the same value of the averaged amplitude $\delta$. The analysis, therefore, can be reduced to the centrally peaked values without loss of generality.} (when $\alpha>1$, meaning the central value is smaller than the peak) or infinite - a problem which is avoided by using the smoothed value (see appendix \ref{AppendixB}). 

\item {\bf The choice of window function has a significant effect on the calculated abundance of PBHs} (as was discussed in \cite{Ando:2018qdb}). The threshold value for collapse is typically stated in terms of the volume-averaged density contrast (as for example in \cite{Musco:2004ak,Musco:2018rwt}), which corresponds to a top-hat window function - suggesting a top-hat function should be used. However, in the super-horizon regime, the smoothing function decreases as $k^2$ but perturbations grow as $k^2$ - meaning the top hat window function is typically not efficient enough at removing small-scale perturbations. For this reason a Gaussian window function is often used in the literature. This however has the drawback of changing the perturbation shape, introducing an 
unphysical change in the value of the threshold $\delta_c$. For this reason, we will follow the standard approach of using a top-hat smoothing function in this paper, whilst treating perturbations as if they are still linear at horizon entry, and employ the linear transfer function for sub-horizon perturbations to reduce the effects of small-scale perturbations (although note that the linear transfer function might not very accurate for the large amplitude perturbations required to generate PBHs). 

\item {\bf The scale of a perturbation $r_m$} is best stated in terms of the radius at which the compactness function $C(r)$ is maximised (see 
Section \ref{sec:ic}).
This is different to the previously used definition $r_0$ \cite{Musco:2004ak}, where the scale of the perturbation was defined as the 
radius at which the density contrast becomes negative. As was shown in \cite{Musco:2018rwt} computing the density contrast at $r_0$  
does not give a sensible parameter to compare different shapes. The averaged value of the density contrast evaluated at $r_m$ is 
characterised by the general relation given by \eqref{delta_m}: thus it relates the local value of the density contrast with the smoothed 
averaged value, independently of any possible choice of the curvature profile. For this reason, measuring the amplitude of the perturbation
at $r_m$ is a consistent way to quantify the effect of the curvature profile on the threshold.
     
\end{itemize}

In this paper, the criteria for a perturbation to collapse to form a PBH will be stated in terms of the volume-averaged density perturbation (averaged over a sphere of radius $r_m$, corresponding to a top-hat window function with radius $r_m$ centred on the peak of the perturbation) at the time of horizon reentry where the perturbation is taken to behave linearly up to this point (although this is not assumed in the simulations). The formation criterion, and its effect on the calculated abundance of PBHs obtained is discussed in more detail in \cite{Young:2019osy}.

\section{Numerical results of PBH formation}
\label{sec:results}

\subsection{Numerical scheme}
The calculations made in this paper to calculate the threshold of PBH formation for the different initial $\zeta$-profiles described in Section \ref{sec:ic} 
have been made with the same code as used in \cite{Musco:2004ak,Polnarev:2006aa,Musco:2008hv,Musco:2012au,Musco:2018rwt}. 
This has been fully described previously and therefore we give only a very brief outline of it here. It is an explicit Lagrangian hydrodynamics 
code with the grid designed for calculations in an  expanding cosmological background. The basic grid uses logarithmic spacing in a mass-type comoving 
coordinate, allowing it to reach  out to very large radii while giving finer resolution at small radii. The initial data follow from the initial condition seen in 
Section \ref{sec:ic},  specified on a space-like slice at constant initial cosmic time $t_i$ defined as $a(t_i)r_m e^{\zeta(r_m) }= 10/H$, ($\epsilon = 10^{-1}$),  
while the outer edge of the grid has been placed at $90 R_m$, to ensure that there is no causal contact between it and the perturbed  region during the 
time of the calculations. The initial data is then evolved using the Misner-Sharp-Hernandez equations so as to generate  a second set of initial data on an 
initial null slice which is then evolved using the Hernandez-Misner equations. During the evolution the grid is modified with an adaptive mesh refinement 
scheme (AMR), built on top of the initial logarithmic grid, to provide sufficient resolution to follow black hole formation down to extremely small values of 
($\delta-\delta_c$).

\subsection{Threshold, scaling law and mass spectrum}
\begin{figure*}[t!]
\vspace{-1.5cm}
 \centering
  \includegraphics[width=0.49\textwidth]{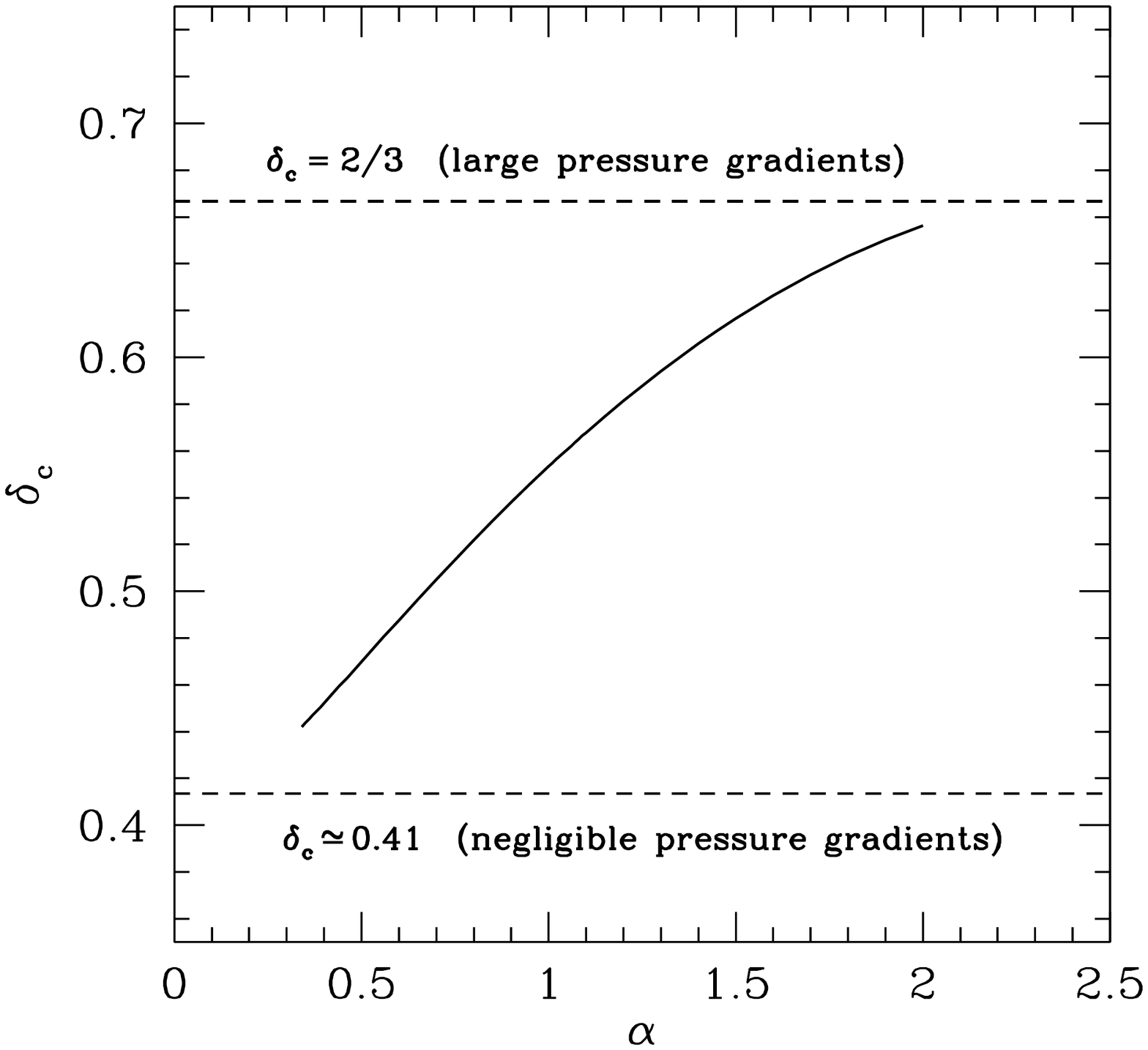} 
  \includegraphics[width=0.49\textwidth]{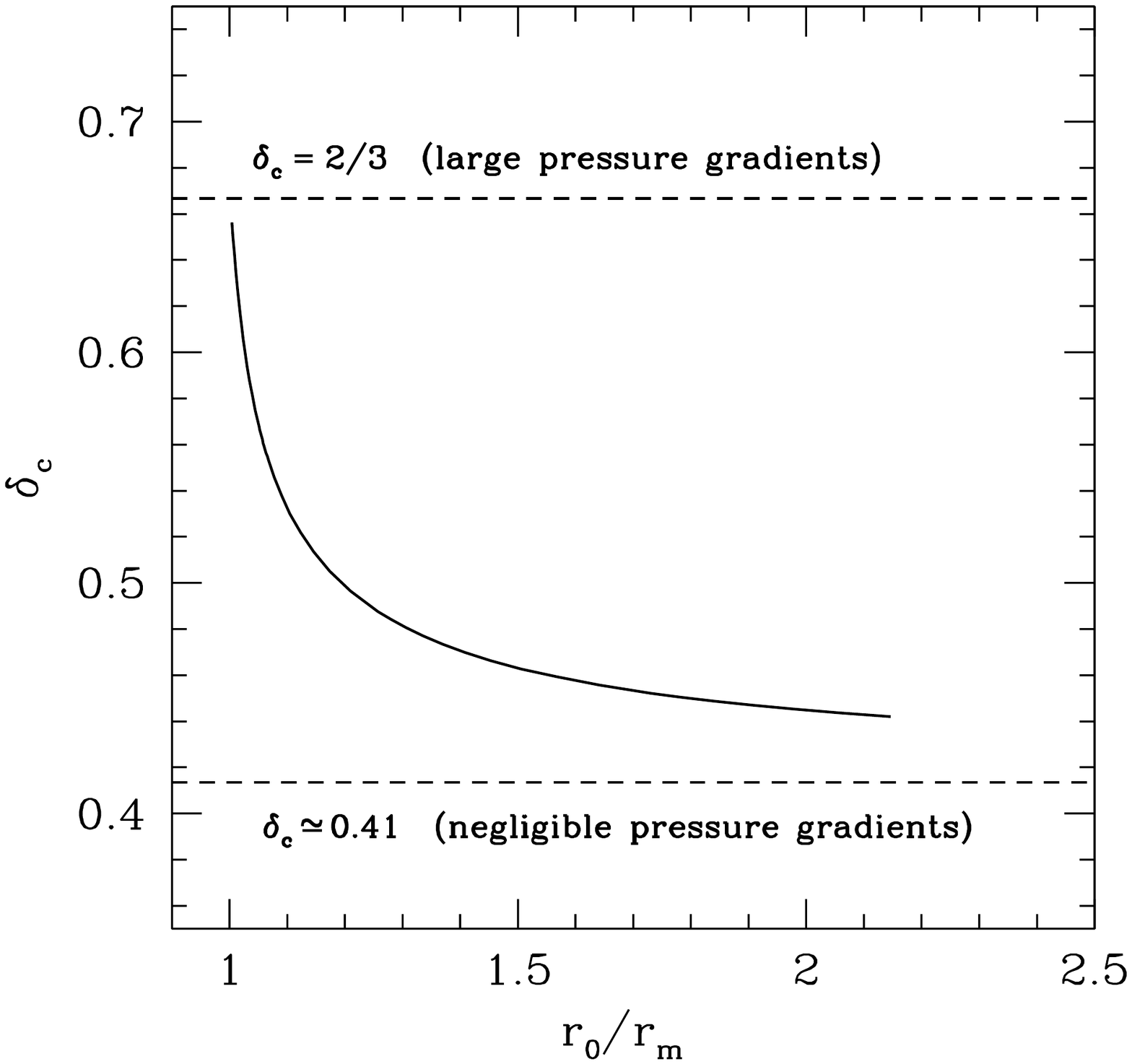} 
  \vspace{-2.0cm}
  \caption{ The left plot show the value of $\delta_c$ as function of $\alpha$ obtained with numerical simulations using an initial curvature profile given by 
  \eqref{zeta_Gauss}, while the right panel show the behaviour of $\delta_c$ as function of the corresponding behaviour if $r_0/r_m$. 
  The bottom dashed horizontal line indicates the lowest limit of the threshold, obtained analytically assuming that the pressure gradients during the 
  collapse are negligible. The upper dashed horizontal shows to the largest possible value of $\delta_c$, with shapes characterized  by very large pressure 
  gradients at the scale $r_m$. } 
  \label{delta_c}
\end{figure*}

In the left panel of Figure \ref{delta_c} we plot the threshold values $\delta_c$ against the parameter $\alpha$ used in  \eqref{zeta_Gauss} to vary the 
initial curvature profile. The lowest limit $\delta_c\simeq0.41$ corresponds to the analytic solution derived in \cite{Harada:2013epa} where the gravitational 
effect of pressure was taken into account while pressure gradients were instead neglected. It was shown in \cite{Musco:2018rwt} that this corresponds to 
shapes of the density contrast with a very large peak ($\alpha\ll1$) and a smooth tail ($r_0 \gg r_m$): in this configuration most of the matter is already within 
in the initial cosmological horizon, and only a negligible amount of matter is spread away by the pressure gradients before the black hole is formed, without 
modifying the shape significantly during the collapse. The maximum value of  $\delta_c=2/3$, ($-r_m\zeta'(r_m)=1$), corresponds instead to shapes with 
$r_m = r_0$, ($\alpha\to\infty$), and the density contrast is very steep. In this case the pressure gradients are very large and a substantial modification of the 
matter configuration is produced during the collapse. 

The right panel show the same results of $\delta_c$ as a function of the corresponding behaviour of $r_0/r_m$: the range of values we have been able to 
compute here are $0.442\lesssim\delta_c\lesssim0.656$, ($0.34\leq\alpha\leq2$). Beyond this range the initial profile of the density contrast becomes too 
extreme, making the numerical simulations unstable. A more detailed analysis of the relationship between the density contrast and the morphology of the initial 
curvature profile can be found in \cite{Musco:2018rwt} where different parameterizations of the curvature profiles has been considered, using more than one 
parameter, getting much closer to the lower limit of $\delta_c\simeq0.41$. 

As has been shown in previous works \cite{Niemeyer:1997mt,Niemeyer:1999ak,Musco:2004ak,Musco:2008hv,Musco:2012au} the mass spectrum of PBHs 
follows the critical collapse, characterized by a scaling law relation
\begin{equation}
\frac{M_{PBH}}{M_H} = {\cal K} (\delta-\delta_c)^\gamma
\end{equation}
where $M_H$ is the mass of the cosmological horizon at horizon crossing, $\gamma\simeq0.36$ is the critical exponent depending only on the value of $\omega$ 
of the equation of state and $\cal{K}$ is a parameter depending on the particular shape of the density contrast. Because this parameter will play some role in the 
next sectionto determine the abundance of PBHs, we have performed numerical simulations to quantify its variation, finding that it varies between $3$ and $11$  
for $\alpha$ varying from $0.4$ and $1.9$  (corresponding to $\delta_c$ varying between $0.45$ and $0.65$) for the profiles considered here, with a representative value of ${\cal K}\simeq4$ when $\delta_c\simeq0.55$, ($\alpha=1$ in \eqref{zeta_Gauss}). 

\section{Calculation of PBH abundance}
\label{sec:abundance}

In this section we will discuss how the abundance of PBHs can be calculated from the primordial curvature perturbation power spectrum $\mathcal{P}_\zeta$. The abundance of PBHs will be stated in terms of the energy fraction of the universe (which will be) contained in PBHs at the time of formation, taken for simplicity to be the time of horizon reentry. In principle the time taken for a PBH to form depends slightly on the amplitude of the perturbation collapsing. A formalism for deriving the mass function from a given power spectrum $\mathcal{P}_\zeta$ taking into account the non-linear relation between $\zeta$ and $\delta_m$ will also be derived.

We will assume throughout that the curvature perturbation has a Gaussian distribution, partly for simplicity and also motivated by the fact that any local-type non-Gaussianity with \mbox{$f_{NL}\gtrsim \mathcal{O}(10^{-3})$} will generate an unacceptably large dark-matter isocurvature perturbation in the CMB \cite{Young:2015kda,Tada:2015noa} - although such bounds can be evaded if non-Gaussianity only couples scales smaller than those observable in the CMB or LSS. In addition, we note that the non-Gaussianity present in single-field inflation (e.g.~the Maldacena consistency relation \cite{Maldacena:2002vr}) does not generate isocurvature perturbations \cite{Pajer:2013ana,Cabass:2016cgp}.  Even when taking ultra slow roll inflation into account, it remains uncertain whether the non-Gaussianity can have a relevant effect \cite{Bravo:2017wyw,Bravo:2017gct,Atal:2018neu,Passaglia:2018ixg} unless the inflaton field has a non-canonical kinetic term \cite{Shandera:2012ke,Young:2015cyn,Franciolini:2018vbk,Kamenshchik:2018sig}.

The density contrast $\delta\rho/\rho_b$ is related to the curvature perturbation $\zeta$ as in equation \eqref{eqn:non-linear}. However, the key parameter 
to use for PBH formation is instead the smoothed density contrast $\delta_m$. Using a top-hat window function with areal radius 
$R=a(t)\exp(\zeta(r_m))r_m$, the amplitude of (spherically symmetric) peaks in the smoothed density contrast is related to the curvature perturbation in radiation domination as \cite{Musco:2018rwt}
\begin{equation}
\delta_m = - \frac{2}{3} r_m \zeta'(r_m)\left[ 2 + r_m\zeta'(r_m) \right].
\label{eqn:smoothNonLinear}
\end{equation} 
Because $\zeta$ has a Gaussian distribution, its derivative will also have a Gaussian distribution. Therefore, equation \eqref{eqn:smoothNonLinear} can be expressed in terms of a linear Gaussian component $\delta_l = -\frac{4}{3}r_m\zeta'(r_m)$ as\footnote{It is noted that a recent paper \cite{Kawasaki:2019mbl} performed a more detailed analysis derived from the skewness of the distribution to achieve the same result (as can be seen from combining equation (30) and (31) in that paper). The correspondence of $-\frac{4}{3}r_m\zeta'(r_m)$ to the linear component of $\delta_r$ and the derivation of its variance are discussed in more detail in Appendix \ref{AppendixA}.}
\begin{equation}
\delta_m = \delta_l - \frac{3}{8}\delta_l^2.
\label{eqn:NL}
\end{equation}
The probability density function (PDF) of $\delta_l$ then follows a Gaussian distribution
\begin{equation}
P(\delta_l) = \frac{1}{\sqrt{2\pi\sigma^2 }}\exp \left( -\frac{\delta_l^2}{2\sigma^2} \right).
\label{eqn:gaussianPDF}
\end{equation}
$\delta_l$ represents the linear component of the smoothed density contrast and its variance $\sigma^2$ can be calculated by integrating the linear component of the density power spectrum as follows:
\begin{equation}
\sigma^2 = \langle \delta_l^2 \rangle = \int\limits_0^\infty \frac{\mathrm{d}k}{k} \mathcal{P}_{\delta_l}(k, r_m) = \frac{16}{81}\int\limits_0^\infty \frac{\mathrm{d}k}{k}(k r_m)^4 \tilde{W}^2(k, r_m) T^2 (k, r_m) \mathcal{P}_\zeta(k),
\label{eqn:variance}
\end{equation}
where $\tilde{W}(k,r_m)$ is the Fourier transform of the top-hat smoothing function, $T(k,r_m)$ is the linear transfer function, and the smoothing scale $r_m$ is equal to the horizon scale. The horizon scale $r_m$ is used to define the time at which 
$\mathcal{P}_{\delta l}$ (and $T(k,r_m)$) should be evaluated\footnote{We note, whilst the initial conditions of the perturbations are defined in the super-horizon regime, as in \eqref{eqn:non-linear}, we will follow the standard approach used in the literature and evaluate PBH abundance at the time a perturbation enters the horizon (as is done in \cite{Kawasaki:2019mbl} for example), although note this this is somewhat \emph{ad hoc}. A recent paper \cite{Kalaja:2019uju} addressed this point more directly, and estimated the effect of a non-linear transfer function on the density contrast at horizon crossing. Further investigation of the effect of the loss of accuracy from using the linear transfer function is left for future work}. 
For simplicity here, we assume that the relevant scale for PBH formation is given by $k r_m\simeq1$, although this is not a very accurate approximation, and the exact relation between $k$ and $r_m$ depends on the profile of the density contrast, which depends on the shape of the power spectrum \cite{Germani:2018jgr}.

The Fourier transform of the top-hat smoothing function is given by
\begin{equation}
\tilde{W}(k,r_m) = 3 \frac{\mathrm{sin}(k r_m)- k r_m \mathrm{cos}(k r_m)}{(k r_m)^3},
\label{eqn:window}
\end{equation}
and the linear transfer function, where we consider $r_m$ as a time dependent measure of the horizon, is given by \cite{Josan:2009qn}
\begin{equation}
T(k,r_m) = 3 \frac{\mathrm{sin}(\frac{k r_m}{\sqrt{3}})- \frac{k r_m}{\sqrt{3}} \mathrm{cos}(\frac{k r_m}{\sqrt{3}})}{(\frac{k r_m}{\sqrt{3}})^3}.
\label{eq:T}
\end{equation}

The most straight-forward method to calculate the abundance of PBHs from a non-Gaussian distribution is to work instead with the Gaussian component of the perturbations \cite{Byrnes:2012yx,Young:2013oia}. To this end, equation \eqref{eqn:NL} is solved with $\delta_R=\delta_c$ to find the critical amplitude of the linear component $\delta_{c,l\pm}$,
\begin{equation}
\delta_{c,l\pm} = \frac{4}{3}\left( 1 \pm \sqrt{\frac{2-3\delta_c}{2}} \right),
\label{eq:delta-pm}
\end{equation}
where there are 2 solutions because equation \eqref{eqn:NL} is quadratic. When necessary, we will take $\delta_c=0.55$ in this paper, the critical value of the volume averaged density contrast when $\zeta$ is taken to have a Gaussian profile shape. However, we note that only the first solution $\delta_{c,l-}$ corresponds to a physical solution - the second solution corresponds to type 2 perturbations. We will therefore take that a PBH will form in regions where $\delta_c < \delta < 2/3$, where $2/3$ is the maximum value for the density contrast given by equation \eqref{eqn:NL}. This corresponds to $\delta_{c,l-}<\delta_l<4/3$.


The final mass of a PBH, $M_{{\rm PBH}}$, which forms during radiation domination depends on the shape and amplitude $\delta_R$ of the initial perturbation, and the horizon mass at horizon reentry $M_{\rm H}$,
\begin{equation}
M_{{\rm PBH}} = \mathcal{K} M_{\rm H} \left( \delta_m - \delta_c \right)^\gamma = \mathcal{K} M_{\rm H} \left( \left[ \delta_l-\frac{3}{8}\delta_l^2 \right] - \delta_c \right)^\gamma,
\label{eqn:PBHmass}
\end{equation}
where $\mathcal{K}$ depends on the profile shape and typically takes a value between 3 and 5.
During radiation domination $\gamma\simeq0.36$, which is the value we will take \cite{Evans:1994pj}. The horizon mass $M_{\rm H}$ is proportional to the horizon scale squared $r_m^2$,
\begin{equation}
\frac{M_{\rm H}}{M_\odot}\approx \left( \frac{r_m}{r_\odot} \right)^2 =  \left( \frac{k}{k_\odot}\right) ^{-2}.
\label{eqn:massScale}
\end{equation}

For a random Gaussian field, the number density of sufficiently rare peaks in a comoving volume is \cite{Bardeen:1985tr}
\begin{equation}
\mathcal{N} = \frac{\mu^3}{4 \pi^2 \sigma^3} \nu^3 \exp \left( -\frac{\nu^2}{2} \right),
\label{eqn:peakdensity}
\end{equation}
where $\sigma$ is given by equation \eqref{eqn:variance}, 
\begin{equation} \nu\equiv\delta_l/\sigma
\label{eqn:nu}
\end{equation} 
and $\mu$ is given by
\begin{equation}
\mu^2 = \int\limits_0^\infty \frac{\mathrm{d}k}{k} \mathcal{P}_{\delta_l}(k, r_m) k^2 = \frac{16}{81}\int\limits_0^\infty \frac{\mathrm{d}k}{k}(k r_m)^4 \tilde{W}^2(k, r_m) T^2 (k, r_m) \mathcal{P}_\zeta(k) k^2.
\label{eqn:firstMoment}
\end{equation}
Here, the application of equation \eqref{eqn:peakdensity} relies on the assumption that peaks in the smoothed linear density field correspond to peaks in the smoothed non-linear density field, such that equation \eqref{eqn:smoothNonLinear} can be applied to calculate the height of peaks in the non-linear field. In general, this will not be true, but will be valid in the case that only sufficiently rare and large peaks are considered (which is the same assumption required for the validity of \eqref{eqn:peakdensity} and spherical symmetry) - discussed further in appendix \ref{AppendixB}.

The fraction of the energy of the universe at a peak of given height $\nu$ which collapses to form PBHs, $\beta_\nu$, then depends on the mass of the PBHs relative to the horizon scale and the number density of the peaks:
\begin{equation}
\beta_\nu = \frac{M_{{\rm PBH}}(\nu)}{M_{\rm H}(r_m)}\mathcal{N}(\nu)\theta(\nu - \nu_c),
\label{beta-first-time}
\end{equation}
where the time dependance of the horizon mass $M_{\rm H}(r_m)$ is parameterised by the horizon scale $r_m$. $\theta(\nu-\nu_c)$ is the Heaviside step function which indicates that no PBH will form if $\nu$ is below the critical threshold. The total energy fraction of the universe contained within PBHs at a single time of formation is given by integrating over the range of $\nu$ which forms PBHs,
\begin{equation}
\beta = \int\limits_{\nu_{c-}}^{\frac{4}{3\sigma}} \mathrm{d}\nu \frac{\mathcal{K}}{3\pi} \left( \nu\sigma-\frac{3}{8}(\nu\sigma)^2 - \delta_c \right)^\gamma  \left( \frac{\mu}{aH \sigma} \right)^3 \nu^3 \exp \left( -\frac{\nu^2}{2} \right),
\label{eqn:beta}
\end{equation}
where $\nu_{c-}\equiv\delta_{c,l-}/\sigma$, see equation \eqref{eq:delta-pm}, and equations \eqref{eqn:PBHmass}, \eqref{eqn:peakdensity}, and \eqref{eqn:nu} have been explicitly substituted into \eqref{beta-first-time}. The total energy fraction of the universe contained within PBHs today can be approximated by integrating over all times at which PBHs form, parameterised here by the horizon mass (more details of this integral can be found in \cite{Byrnes:2018clq})
\begin{equation}
\Omega_{{\rm PBH}} = \int\limits_{M_{\rm min}}^{M_{\rm max}}\mathrm{d} (\mathrm{ln}M_{\rm H}) \left( \frac{M_{\rm eq}}{M_{\rm H}} \right)^{1/2}\beta(M_{\rm H}),
\label{eqn:omega}
\end{equation}
where $M_{\rm H}$ is the horizon mass at the time of formation and $M_{{\rm eq}}$ is the horizon mass at the time of matter-radiation equality. We will use the value $M_{{\rm eq}}=2.8\times10^{17}M_\odot$ (using the same parameters as \cite{Nakama:2016gzw}). $M_{\rm min}$ and $M_{\rm max}$ are respectively the smallest and largest horizon masses at which PBHs form\footnote{In principle $M_{\rm max}$ can be arbitrarily large because $\delta_m-\delta_c$ can be arbitrarily small. However, the largest PBH mass which can be formed when the horizon mass is $M_{\rm H}$ is given by $M_{PBH} = \mathcal{K}M_{\rm H} \left( 2/3 - \delta_c \right)^\gamma$.}. The derivation of this formula assumes that the matter density $\Omega_m$ during radiation domination grows proportionally to the scale factor until the time of matter-radiation equality, whereupon the universe immediately becomes matter dominated.

The mass function $f(M_{{\rm PBH}})$ can then be obtained by differentiating $\Omega_{{\rm PBH}}$ with respect to the PBH mass:
\begin{equation}
f(M_{{\rm PBH}}) = \frac{1}{\Omega_{CDM}} \frac{\mathrm{d}\Omega_{{\rm PBH}}}{\mathrm{d}(\mathrm{ln}M_{{\rm PBH}})},
\label{eqn:massFunction}
\end{equation}
where the equations \eqref{eqn:variance}, \eqref{eqn:firstMoment} and \eqref{eqn:beta} should be recast in terms of the horizon mass and PBH mass using the substitutions in equations \eqref{eqn:PBHmass} and \eqref{eqn:massScale}. $\Omega_{CDM}$ will be taken as 0.27 where necessary in this paper.

Broad and narrow peaks in the power spectrum are often considered when calculating the PBH abundance. To give a concrete example, we will consider the two extreme cases - a scale invariant power spectrum, and an extremely narrow peak. For the scale invariant case, we will take $\mathcal{P}_\zeta=\mathcal{A}_s=constant$. For the narrow peak, we will take the peak to be narrow enough such that it can be treated as a Dirac-delta function, $\mathcal{P}_\zeta=\mathcal{A}_s \delta_D(\mathrm{ln}(k/k_*))$, although note that such a power spectrum is unphysical (for example, \cite{Byrnes:2018txb} describes that the power spectrum cannot be steeper than $k^4$, at least in the context of single-field inflation).

\begin{figure}
\centering
\includegraphics[width=0.47\textwidth]{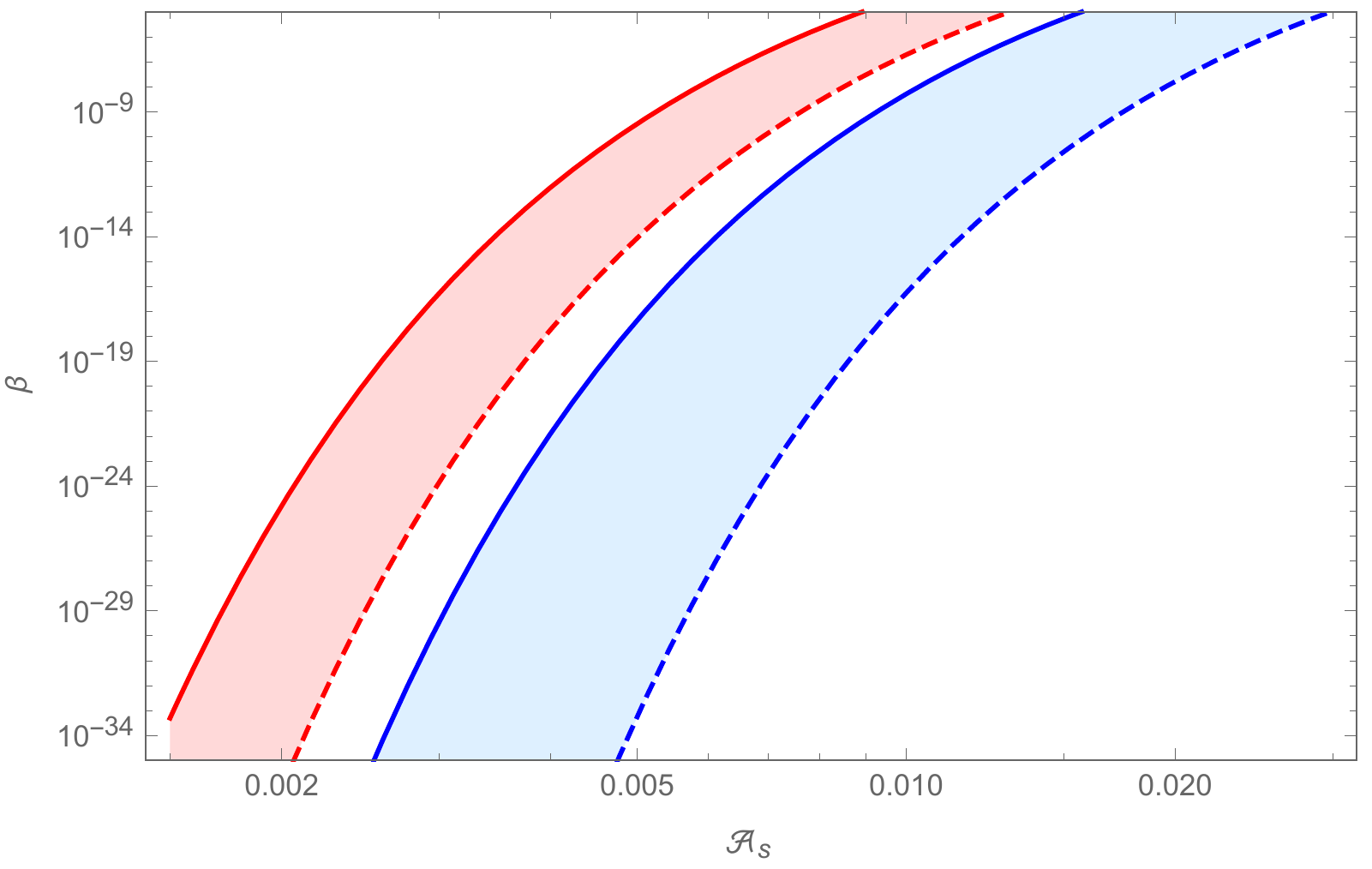} 
\hspace{0.5cm}
\includegraphics[width=0.47\textwidth]{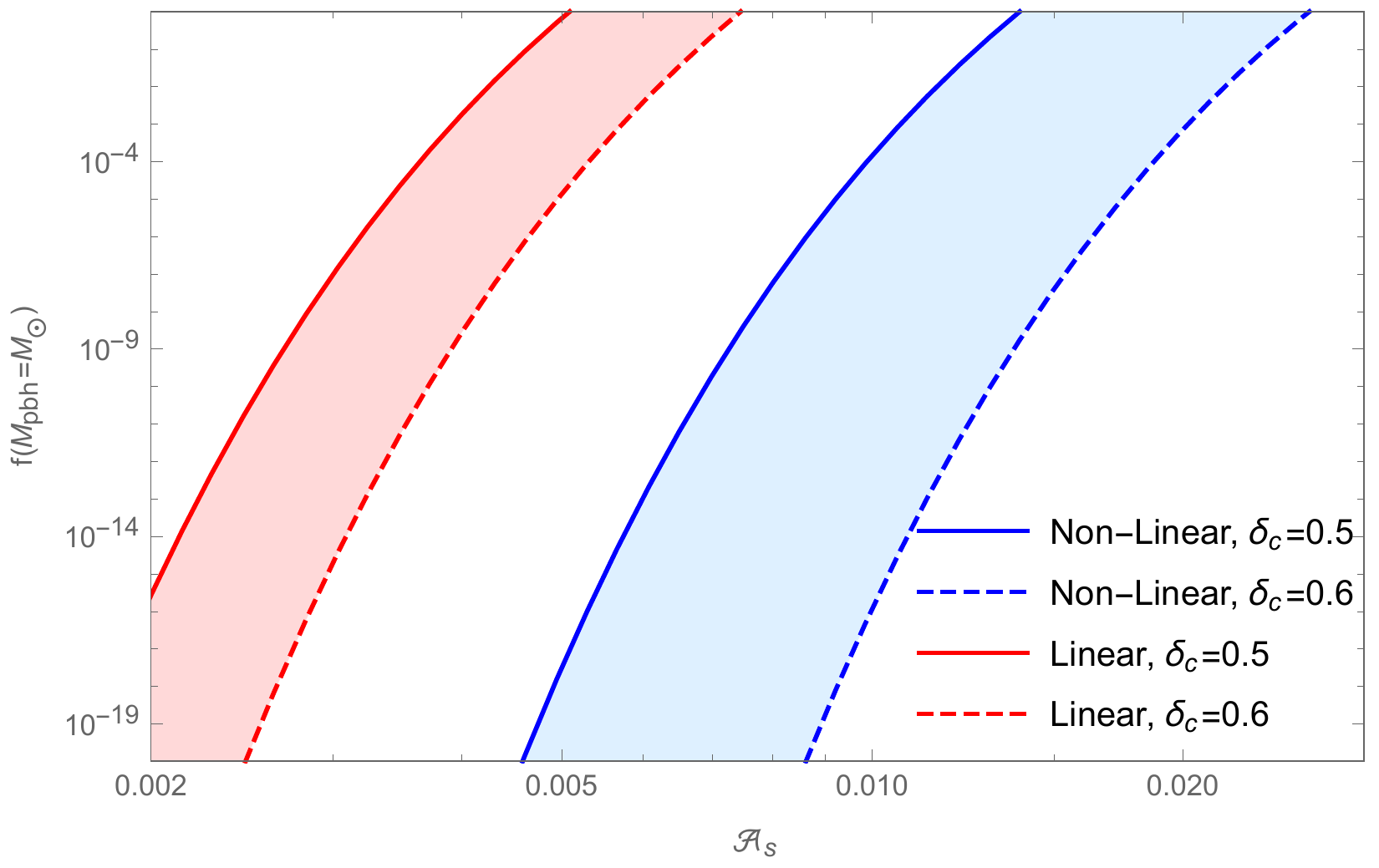} 
\caption{ Top: The energy fraction of the universe collapsing to form PBHs at the time of formation, $\beta$, is plotted against the amplitude of a scale invariant power spectrum, $\mathcal{P}_\zeta=\mathcal{A}_s$. Bottom:  The mass function of PBHs $f(M_\odot)$ of PBHs (defined in equation \eqref{eqn:massFunction}) with a mass of $1 M_\odot$ as a function of the amplitude of $\mathcal{A}_s$. Accounting for the non-linear relation means that a significantly smaller abundance of PBHs is calculated, by many orders of magnitude. For $\delta_c=0.5$ then $K=5.5$ and for $\delta_c=0.6$ then $K=3.5$. The abundance of PBHs is dominated by the uncertainty in $\delta_c$ (which has an exponential effect), rather than $K$ (which has a linear effect).}
\label{fig:BetaVsAs}
\end{figure}

For the scale invariant case, equations \eqref{eqn:variance} and \eqref{eqn:firstMoment} give scale invariant values of $\sigma^2 \sim 1.06 \mathcal{A_s}$ and $(\mu/(aH))^2 \sim 6.86 \mathcal{A}_s$ respectively. Figure \ref{fig:BetaVsAs} shows the abundance of PBHs, $\beta$ (equation \eqref{eqn:beta}), as a function of $\mathcal{A}_s$, whilst figure \ref{fig:BetaVsAs} shows the mass function $f(M)$ (equation \eqref{eqn:massFunction}) evaluated at $M=M_{{\rm PBH}}$ as a function of the power spectrum amplitude. 

In the case of the narrowly peaked spectrum, we assume without loss of generality that the power spectrum peaks at a scale corresponding to a horizon mass of $1M_\odot$, with corresponding horizon scale $k_* = k_\odot$, in order to allow a direct comparison to the broad power spectrum. We will also consider that PBH formation occurs only at the horizon scale corresponding exactly to the peak in the power spectrum, and that $\beta$ corresponds to the total energy fraction collapsing into PBHs at all epochs, rather than integrating over $\mathrm{ln}M_{\rm H}$ as in equation \eqref{eqn:omega}. Whilst equation \eqref{eqn:variance} will give a significant variance $\sigma^2$ for values of $r_m$ close to $r_\odot$ (suggesting perturbations of varying scales exist), this is because $\delta_m$ does not immediately become zero when the smoothing scale is not set exactly equal to the scale of the perturbation. 
However, the scale of all perturbations here is fixed, and so will all enter the horizon at the same scale. In this case, evaluated as $k_\odot$ enters the horizon, equations \eqref{eqn:variance} and \eqref{eqn:firstMoment} give $\sigma^2 = (\mu/(aH))^2  \sim 0.151 \mathcal{A}_s$.

\begin{figure}
 \centering
 \includegraphics[width=0.47\textwidth]{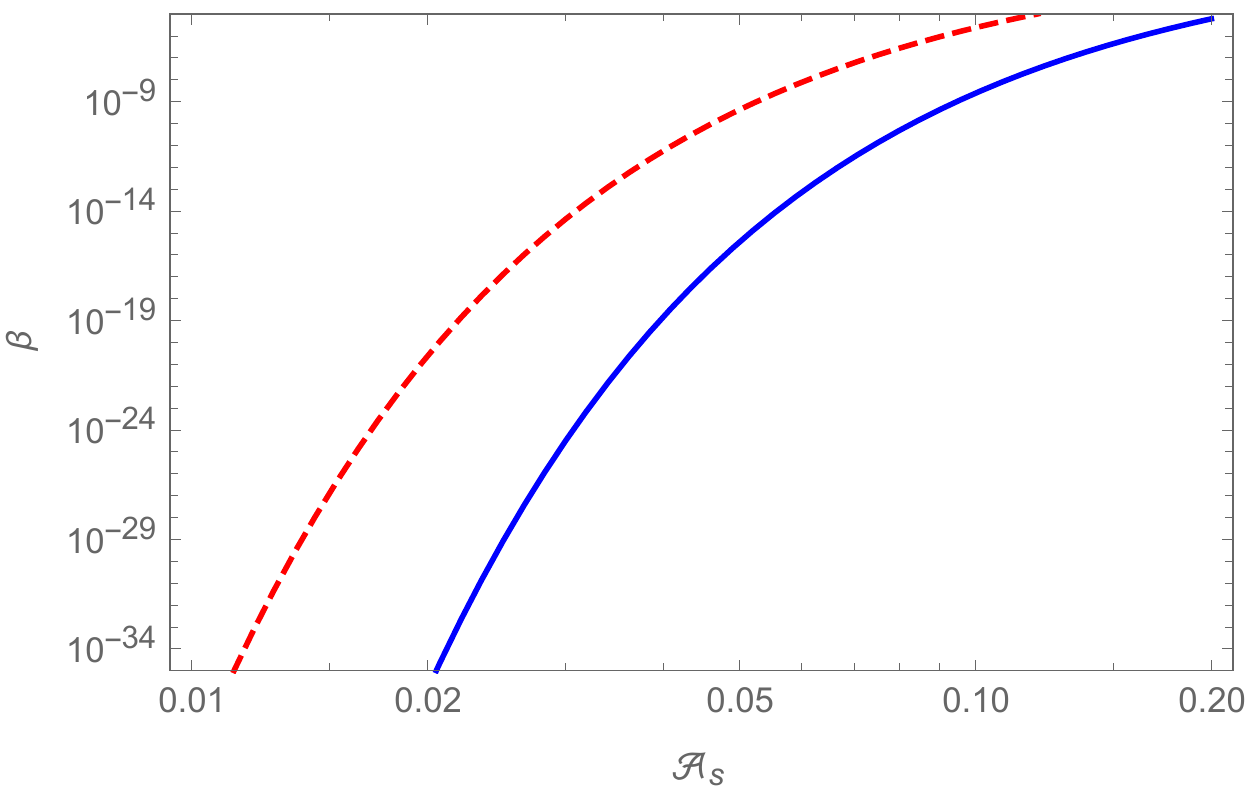} 
\hspace{0.5cm}
 \includegraphics[width=0.47\textwidth]{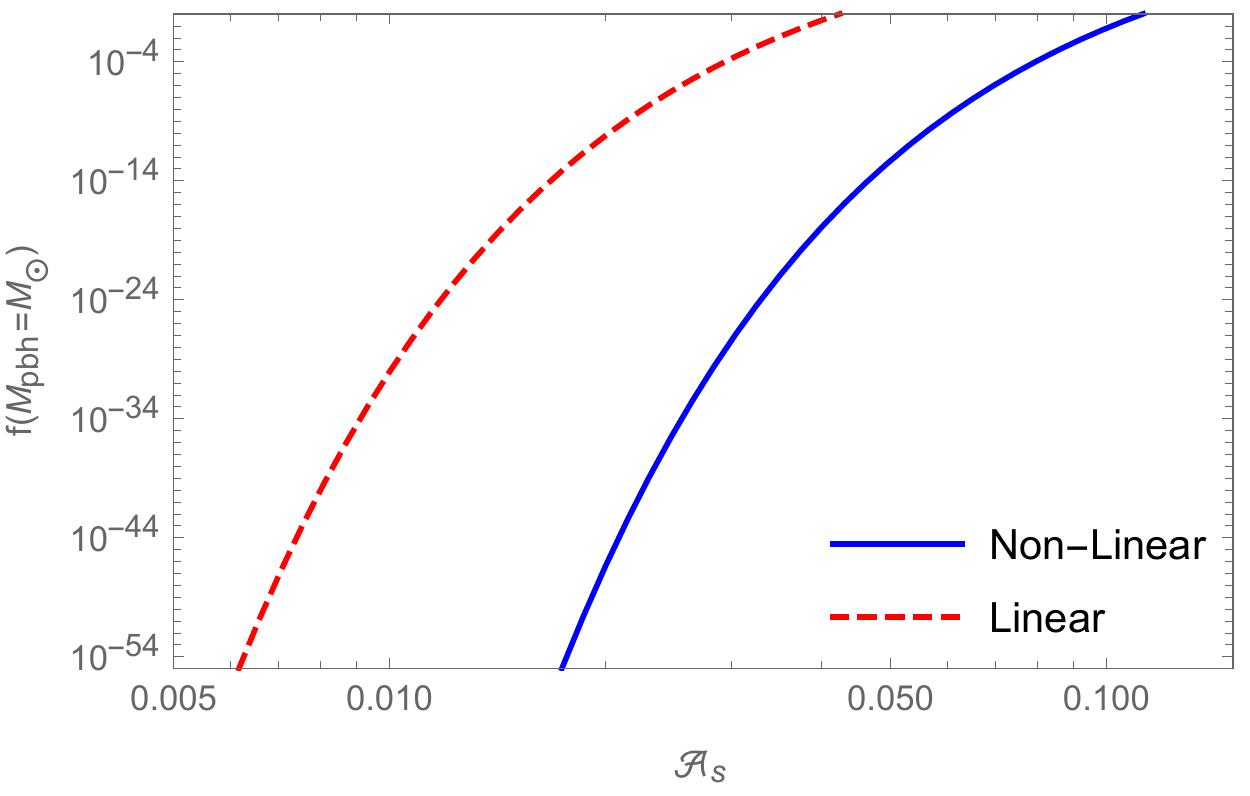} 
  \caption{ Left: The energy fraction of the universe collapsing to form PBHs at the time of formation, $\beta$, is plotted against the amplitude of the power spectrum. Right: the mass function of PBHs $f(M_\odot)$ of PBHs with a mass of $1 M_\odot$ as a function of the amplitude of the power spectrum. In both plots, a Dirac delta power spectrum has been assumed, $\mathcal{P}_\zeta=\mathcal{A}_s \delta_D(\mathrm{ln}(k/k_*))$. Using the profile shape predicted from a Dirac-delta power spectrum, $K=4$, and $\delta_c=0.51$.}
  \label{fig:BetaNarrowPeak}
\end{figure}


What can be learned from these figures is that accounting for the non-linear relationship between the density contrast and the curvature perturbation will always serve to reduce the calculated abundance of PBHs, compared to using the linear model. It can be seen from comparing the figures from the scale invariant power spectrum to the narrowly peaked power spectrum, that the abundance of PBHs is strongly dependent on the shape of the power spectrum rather than simply the amplitude - and so constraints on the power spectrum from constraints on PBH abundance should be calculated on a case-by-case basis for different inflationary models which predict different shapes for the power spectrum. This fact has been well known for some time and was investigated in more detail recently by \cite{Germani:2018jgr}. 

However, we will here make a simple calculation to describe by what fraction the amplitude of the curvature perturbation power spectrum $\mathcal{A}_s$ needs to increase in order to obtain the same value for $\beta$ as when the linear relation is used. The dominant term in the calculation for the abundance of black holes, equation \eqref{eqn:beta}, is the exponential term $\exp(-\nu^2)$. After integrating, this will give a dependance roughly proportional to $\exp(-\nu_{c-}^2)$. In order to produce (approximately) the same number of PBHs from the linear calculation as from the non-linear calculation, we therefore require $\nu_{c,L}=\nu_{c,NL}$ - where the subscripts $L$ and $NL$ denote the linear and non-linear calculations respectively,
\begin{equation}
\frac{\delta_{c,L}}{\sigma_L} = \frac{\delta_{c,l-}}{\sigma_{NL}},
\end{equation}
where $\delta_{c,l-}$ is given by equation \eqref{eq:delta-pm}. In both the linear and non-linear case, $\sigma^2$ for a given power spectrum shape is proportional to $\mathcal{A}_s$ by the same constant of proportionality, $\sigma^2 = C \mathcal{A}_s$. Finally, we can write down the ratio between $\mathcal{A}_L$ and $\mathcal{A}_{NL}$ as
\begin{equation}
\frac{\mathcal{A}_{NL} }{\mathcal{A}_{L}}=\left( \frac{\delta_{c,l-}}{\delta_{c}} \right)^2=\left( \frac{4\left( 1 - \sqrt{\frac{2-3\delta_c}{2}} \right)}{3\delta_c} \right)^2.
\label{eq:Aratio}
\end{equation}
For values $0.41<\delta_c<2/3$, this factor varies approximately from 1.5 to 4. For more typical values $0.5<\delta_c<0.6$, in order for the same number of PBHs to form, the amplitude of the power spectrum must be 1.78--2.31 times greater than previously calculated with the linear model. In particular, for the commonly considered case of a Gaussian profile in $\zeta$, with $\delta_c=0.55$, the power spectrum needs to be enhanced by a factor of 2.0 in order to get the same number of PBHs forming if one neglected the non-linear relationship between $\zeta$ and $\delta$.






\section{Summary}
\label{sec:discussion}

In the radiation dominated epoch following the end of inflation, 
perturbations can collapse to form PBHs if the perturbation amplitude is large enough. Whether or not a PBH will form depends on the amplitude of the density contrast $\delta_m$, rather than the amplitude of the curvature perturbation $\zeta$. The non-linear relation between the curvature perturbation $\zeta$ and the density contrast $\delta\rho/\rho_b$, given by equation \eqref{eqn:non-linear}, means that $\delta_m$ will have a non-Gaussian distribution even when $\zeta$ is perfectly Gaussian, which has a significant effect on the number of PBHs which form in the early universe.

We have discussed briefly the formation criterion which should be used to determine whether a perturbation will collapse or not, and in this paper, we argue that the volume-averaged (smoothed) density contrast of a peak $\delta_m$ should be used rather than the value $\delta\rho/\rho_b$ evaluated at $r=0$. The scale over which the perturbation should be averaged, $r_m$, is the scale at which the compaction function $\mathcal{C}(r)$ (defined in equation \eqref{a}) is maximised, and note that at this scale $\mathcal{C}(r)$ is equal to $\delta_m$. In this paper, we use the amplitude of $\delta_m$ measured at the cosmological horizon entry to determine whether a PBH will form - although point out that this is somewhat inconsistent as the expression for $\delta\rho/\rho_b$ is valid in the super-horizon regime and the perturbation will not evolve linearly until horizon entry when they have a large amplitude. Further consideration of these factors is left for future study.

Making use of numerical simulations described in section \ref{sec:results}, we have considered different profiles of $\zeta$ which form density perturbations to determine a threshold value for PBH formation and how this can depend on the shape of the profile in $\zeta$. We noted in section \ref{sec:ic} that the relation between peaks in $\zeta$ and $\delta\rho/\rho_b$ is not straightforward and may not coincide. For a given profile shape of $\zeta$, the profile of $\delta\rho/\rho_b$ depends also on the amplitude of the perturbation, and we have calculated a relation between the amplitude of the perturbation $\delta_m$ and the mass of the PBH formed 
accounting for this (often referred to as the extended mass function, rather than assuming monochromatic formation of PBHs at a given epoch).

There is a simple relation between $\delta_m$ and $\zeta$, given by equation \eqref{eqn:smoothNonLinear}, which can be used to fully describe the non-Gaussianity of $\delta_m$ when $\zeta$ is taken as Gaussian. Making use of this relation we have derived a formalism to derive the abundance of PBHs. The abundance of PBHs may be expressed either in terms of the energy fraction collapsing into black holes at the time of formation $\beta$, the present-time density parameter for PBHs, $\Omega_{PBH}$, or the mass fraction of dark matter contained within PBHs of a given mass, $f(M_{PBH})$. These expressions are calculated utilising the theory of peaks and accounting for the extended mass function of PBHs.

When the non-Gaussianity of the density $\delta_m$ is taken into account, we find that this always reduces the number of PBHs which form by many orders of magnitude, see figures \ref{fig:BetaVsAs}-\ref{fig:BetaNarrowPeak}. We reproduce the known result that the abundance of PBHs depends upon both the amplitude and shape of the curvature perturbation power spectrum. In order for a comparable number of PBHs to form compared to using the linear relation between $\zeta$ and $\delta$, we find that the amplitude of the power spectrum must therefore be $\sim 2-3$ times larger, with the amount depending on the value taken for the collapse threshold, see \eqref{eq:Aratio}, but otherwise being almost independent of the shape of the power spectrum.

Finally we note that the non-linear relation for the smoothed density contrast in terms of the spatial derivative of the curvature perturbation makes a clear analogy to local non-Gaussianity, with a negative value of (local) $f_{\rm NL}$ that suppresses PBH formation (see also the ``Note added'' below).  However, the analogy is potentially misleading since this non-Gaussian term only affects the one-point function of $\delta_m$ and it does not generate a bispectrum because the derivatives of $\zeta$ are uncorrelated on scales much larger than the PBH scale, $r_m$. Therefore it does not correspond to a coupling between long and short-wavelength modes and hence does not generate a large-scale dark matter isocurvature perturbation (which would be observationally ruled out \cite{Tada:2015noa,Young:2015kda}). This also implies that the ``apparent'' negative local $f_{\rm NL}$ cannot lead to an enhanced formation of PBHs in the case of a primordial power spectrum with a broad peak, in contrast to a physical value of $f_{\rm NL}<0$ of $\zeta$, see \cite{Young:2014oea} for details. 

The relation used between $\zeta$ and $\delta$, equation \eqref{eqn:non-linear}, is exact in $\zeta$, only neglecting higher-order terms in $\epsilon$, valid because $\epsilon\ll 1$ in the super-horizon limit. Therefore, we have quantified the effect of the complete non-linear relationship upon the abundance of PBHs.

{\bf Note added:} While this paper was approaching completion a related work by Kawasaki and Nakatsuka \cite{Kawasaki:2019mbl} appeared on the arXiv. Our broad conclusions are in agreement with their paper, principally that the non-linear relationship between $\zeta$ and $\delta$ makes the formation of PBHs less likely. For typical values of $\delta_c\sim 0.5$, we confirm their finding that the suppression of PBH formation due to the non-linear terms requires the power spectrum to have an amplitude $\sim1.4^2$ times greater in order to form the same number of PBHs.  

Produced and appearing in parallel with our paper, the paper by De Luca et al.~\cite{DeLuca:2019qsy} studies the same topic. Although there are some significant differences in our methodology, our results are in broad agreement, both showing that the non-linear relation between $\delta\rho/\rho_b$ and $\zeta$ leads to a suppression in the PBH formation rate.
Unlike these two papers, we take critical collapse into account.

\section*{Acknowledgements}

We thank V. De Luca, G. Franciolini, A. Kehagias, M. Peloso, A. Riotto and C. \"Unal, the authors of \cite{DeLuca:2019qsy}, for sharing their draft of a paper which was written in parallel to this one and helpful discussions. We thank Nicola Bellomo and Eiichiro Komatsu for helpful comments on a draft of this paper and we thank Jaume Garriga, Cristiano Germani, Marcello Musso and Licia Verde for helpful discussions. 

SY is funded by a Humboldt Research Fellowship for Postdoctoral Researchers. IM is supported by the Unidad de Excelencia Mar{\'i}a de Maeztu Grant No.~MDM-2014-0369. CB is funded by a Royal Society University Research Fellowship.

\appendix

\section{Gaussianity and variance of $-\frac{4}{3}r_m\zeta'(r_m)$}

\label{AppendixA}

Here we will derive the variance and discuss the Gaussianity of the linear term in equation \eqref{eqn:smoothNonLinear}, $\delta_l=-\frac{4}{3}r_m\zeta'(r_m)$. 
The linear relation between the density contrast $\delta\rho/\rho_b$ and the curvature perturbation $\zeta$ in radial coordinates $r$ is given by
\begin{equation}
\frac{\delta\rho}{\rho_b}(r,t) = -\frac{4}{9}\epsilon^2(t) r_m^2 \left( \zeta''(r)+\frac{2}{r}\zeta'(r) \right),
\end{equation}
where the prime denotes a derivative with respect to the radial co-ordinate $r$, and the parameter $\epsilon$ describes the super-horizon time-evolution of the 
perturbation and is given in the linear approximation by $\epsilon=(r_m aH)^{-1}$, where $r_m$ is the perturbation scale and $aH$ is the horizon scale. 
In this paper we consider the perturbations at the time of horizon reentry, $t_H$, such that $r_m=(aH)^{-1} \,  \Rightarrow \, \epsilon(t_H)=1$, and we will therefore drop the $\epsilon(t)$. 
The volume averaged density contrast used to determine PBH formation is given by
\begin{equation}
\delta_m = \frac{1}{V}\int\limits_0^{r_m}  4 \pi r^2 \frac{\delta\rho}{\rho_b}(r,t_H)\mathrm{d}r \,,
\label{eqn:volumeAverage}
\end{equation}
where in the linear approximation the volume is given by $V = \frac{4}{3}\pi r_m^3$. 
Although this integral is normally integrated over the comoving distance $r_{\rm com}\equiv re^{\zeta(r)}$, this difference represents higher-order effects neglected in 
the linear approximation. Combining the above equations gives
\begin{equation}
\delta_m = -\frac{4}{3r_m}\int\limits_0^{r_m} \mathrm{d}r r^2 \left( \zeta''(r)+\frac{2}{r}\zeta'(r) \right) = -\frac{4}{3}r_m \zeta'(r_m) = \delta_l \,,
\end{equation}
which is the linear term seen in equation \eqref{eqn:smoothNonLinear}.

We now consider the variance of $\delta_m$, noticing that equation \eqref{eqn:volumeAverage} can be considered as a top-hat window function centred on the peak convolved with the density contrast
\begin{equation}
\delta_m(X) = \frac{1}{V}\int\limits_0^{\infty} \mathrm{d}r 4 \pi r^2 \frac{\delta\rho}{\rho_b}(r,t_H)\theta(r_m-r) =  \int \mathrm{d}x^3 \frac{\delta\rho}{\rho_b}(x,t_H) W(X-x,r_m) \,,
\label{eqn:convolution}
\end{equation}
where $X$ is the location of a peak in cartesian coordinates (corresponding to $r=0$) and $\theta(r)$ is the Heaviside step function. The second equality is the same integral expressed in cartesian coordinates, with the window function $W(x,r_m)$ given by:
\begin{equation}
W(x,r_m) = \frac{\theta(r_m - \left| x \right|)}{\frac{4}{3}\pi r_m^3} \,.
\end{equation}
For our purposes, it is more convenient to express equation \eqref{eqn:convolution}, a convolution in real space, as a multiplication in Fourier space,
\begin{equation}
\delta_m(k) = \tilde{W}(k,r_m)\delta(k) \,,
\end{equation}
where the Fourier transform of the window function $\tilde{W}(k,r_m)$ is given by equation \eqref{eqn:window} and $\delta(k)$ is given by
\begin{equation}
\delta(k) = -\frac{4}{9}T(k,r_m)(k r_m)^2 \zeta(k) \,.
\end{equation}
where the linear transfer function $T(k,r_m)$ is given by equation \eqref{eq:T} and $\delta_m(k)$ can therefore be written as
\begin{equation}
\delta_m(k) = -\frac{4}{9}\tilde{W}(k,r_m)T(k,r_m)(k r_m)^2\zeta(k) \,.
\end{equation}
Since we have assumed $\zeta(x)$ has a Gaussian distribution, it also has a Gaussian distribution in Fourier space: $\zeta(k)$ has a Gaussian distribution (being a linear combination of Gaussian variables $\zeta(x)$). $\delta_m(k)$ is then related to $\zeta(k)$ by a linear factor, meaning that $\delta_m(k)$ also has a Gaussian distribution. Finally, $\delta_m(x)$ is a linear combination of the Gaussian Fourier modes; hence it also has a Gaussian distribution.

Finally, we can calculate the variance $\sigma^2$ by integrating the power spectrum,
\begin{equation}
\sigma^2 = \langle \delta_l^2 \rangle = \int\limits_0^\infty \frac{\mathrm{d}k}{k} \mathcal{P}_{\delta l}(k, r_m) = \frac{16}{81}\int\limits_0^\infty \frac{\mathrm{d}k}{k}(k r_m)^4 \tilde{W}^2(k, r_m) T^2 (k, r_m) \mathcal{P}_\zeta(k) \,.
\end{equation}

\section{Correspondence of large peaks}

\label{AppendixB}

We will here consider the correspondence between peaks in the various fields considered within the context of this paper - the curvature perturbation field $\zeta$, the non-linear density contrast field $\delta\rho/\rho_b$, the non-linear smoothed density field $\delta_m$ and the linear smoothed density field $\delta_l$. For the purposes of this discussion, type 2 perturbations will not be considered (corresponding to $\delta_m>2/3$, as the abundance of such perturbations is exponentially suppressed (even compared to the exponentially small number of perturbations which form PBHs).

\begin{figure}
\centering
\includegraphics[width=0.7\textwidth]{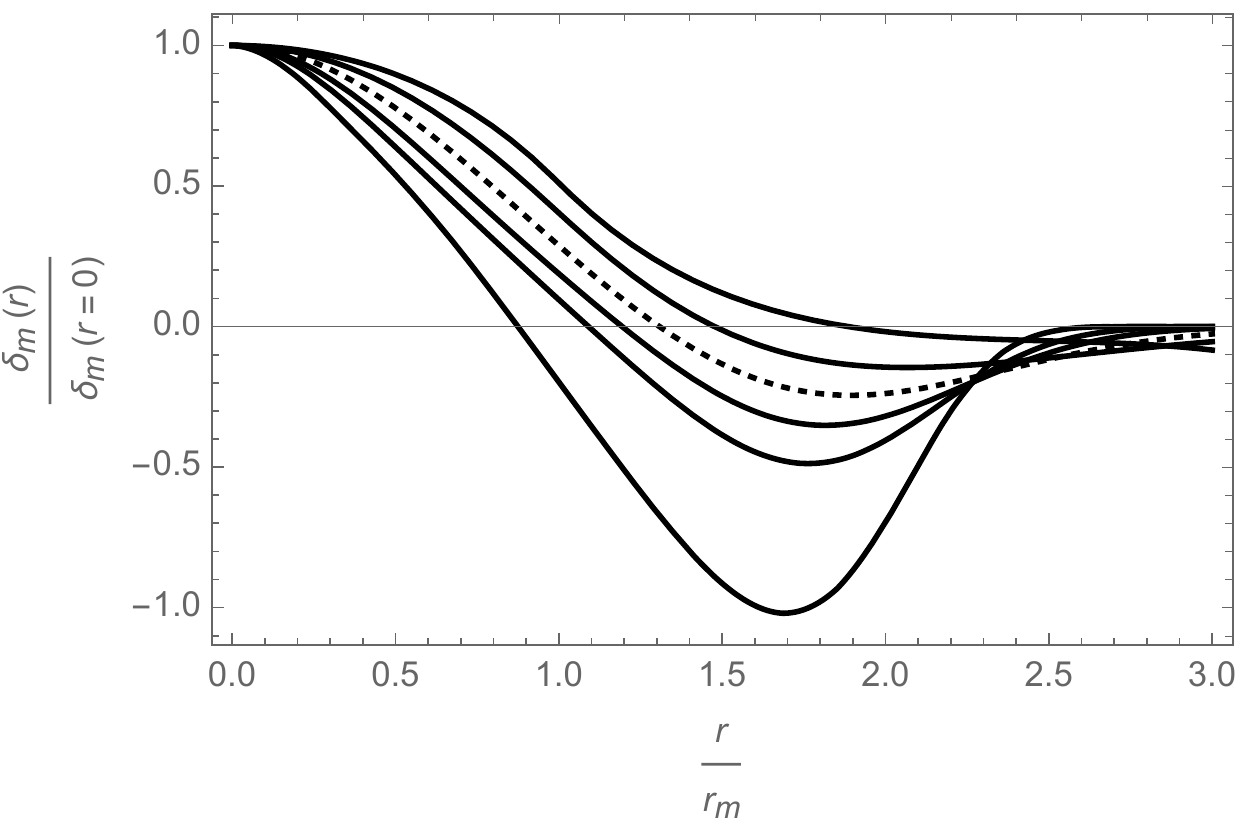} 
\caption{ Profile shapes in the density contrast after a top-hat smoothing function has been applied. The profiles are calculated from a curvature perturbation profile given by equation \eqref{zeta_Gauss}, and each profile is smoothed on a scale $r_m$. The values of $\alpha$ are $\alpha=0.5, 0.75, 1.0, 1.25, 1.50, 2.0$, where increasing $\alpha$ corresponds to more negative troughs seen in the figure. The amplitude in each case is $\mathcal{A}=0.5$.}
\label{fig:smoothedprofiles}
\end{figure}

Figure \ref{delta_rho} shows the critical profiles of peaks in $\zeta$ and $\delta\rho/\rho_b$ (in spherical symmetry). All of the profiles in $\zeta$ have a central peak at $r=0$, whilst the density field can have an off-centred peak or a divergence at the centre - and it cannot necessarily therefore be stated that a PBH forms at peaks in $\delta\rho/\rho_b$. As mentioned in section \ref{sec:collapse} this problem is overcome by using the smoothed density contrast - figure \ref{fig:smoothedprofiles} shows the same profiles smoothed on a scale $r_m$. The off-centred peaks are no longer seen because the smoothing scale $r_m$ will by definition larger than the radius at which the density peaks, and, being a feature smaller than the smoothing radius, is therefore removed in the process of smoothing. The divergences at the centre (while unphysical) are nonetheless removed because the divergence, whilst they represent infinite density as $r\rightarrow 0$, they do not represent infinite mass - and the integral during the smoothing therefore converges. It can therefore be stated that, for isolated spherically-symmetric type 1 perturbations, the peaks in $\zeta$ correspond to peaks in the smoothed density field $\delta_m$.

In order to calculate the abundance, it has been assumed that peaks in the (Gaussian) linear smoothed density field correspond to peaks in the (non-Gaussian) non-linear field. In general, this is not expected to be true - however, for the very large and rare peaks from which PBHs form, it is expected that this should be a valid approximation. This is primarily due to the fact that the large peaks in the Gaussian fields ($\zeta$ or $\delta_l$) are very unlikely to be close to other large peaks, that is, that the local region surrouding large peaks contains only significantly smaller perturbations. This is expected to result in the large peak being (approximately) spherically symmetric peak \cite{Bardeen:1985tr}. When the non-linear smoothed density field is calculated, this spherical symmetry is preserved - implying that the perturbation in the non-linear field must also be a peak (when smoothed on an appropriate scale, see above). 

The correlation of peaks in $\zeta$ and the density $\delta\rho/\rho_b$ was investigated in \cite{DeLuca:2019qsy}, concluding that \textit{``one can associate the number of rare peaks in the overdensity with the number of peaks in the curvature perturbation which are spiky enough''}, validating the assumptions applied here.

Furthermore, only modes which are similar to the smoothing scale (which is considered equal to the horizon scale) have a non-negligible effect on the smoothed density field - larger-scale and smaller-scale modes are suppressed by the $k^4$ term and the smoothing and transfer functions respectively in equation \eqref{eqn:variance}. The smoothed density fields (linear and non-linear) therefore only inherit peaks from the $\zeta$-field on a small range of scales. This means that, whilst a perturbation of scale $r_m$ in the $\zeta$-field is likely to have many scaller-scale peaks on top of it (and may also sit on top of a much larger scale peak), this is not the case for the smoothed density fields.

\bibliographystyle{JHEP} 
\bibliography{bibfile.bib}

\end{document}